\documentclass[a4paper,11pt]{article}
\usepackage{jcappub} % for details on the use of the package, please see the JINST-author-manual
\usepackage{subcaption}
\usepackage{lineno}
\title{\boldmath Quasinormal Modes, Grebody Factors, and Hawking Radiation Sparsity of Black Holes Influenced by a Global Monopole Charge in Kalb-Ramond Gravity}

\author[a]{A. Baruah}
\author[b,c]{Y. Sekhmani}
\author[d]{S. K. Maurya}
\author[a]{A. Deshamukhya}
\author[d]{M. K. Jasim}
\affiliation[a]{Department of Physics, Albert Einstein School of Physical Sciences, Assam University, Silchar - 788011, Assam, India}
\affiliation[b]{Center for Theoretical Physics, Khazar University, 41 Mehseti Street, Baku, AZ1096, Azerbaijan.}
\affiliation[c]{Centre for Research Impact \& Outcome, Chitkara University Institute of Engineering and Technology, Chitkara University, Rajpura, 140401, Punjab, India}
\affiliation[d]{Department of Mathematical and Physical Sciences, College of Arts and Sciences, University of Nizwa, Nizwa 616, Sultanate of Oman}
\emailAdd{anshuman.baruah@aus.ac.in}
\emailAdd{sekhmaniyassine@gmail.com}
\emailAdd{sunil@unizwa.edu.om}
\emailAdd{atri.deshamukhya@aus.ac.in}
\emailAdd{mahmoodkhalid@unizwa.edu.com}

\abstract{Kalb-Ramond (KR) gravity is an intriguing model incorporating local Lorentz violation, and black hole (BH) solutions are known to exist. In this study, we investigate some crucial aspects of BHs endowed with a global monopole charge in the self-interacting KR field. Specifically, we study the quasinormal modes (QNMs) corresponding to scalar, electromagnetic, and gravitational perturbations; derive rigorous bounds for the greybody factors (GBFs); and examine the sparsity of Hawking radiation. The effects of the model parameters $\ell$ (Lorentz-violating parameter in KR gravity) and $\eta$ (monopole charge) on these phenomena are elaborated. First, QNMs are evaluated with high precision using the 13\textsuperscript{th}-order Pad\'{e}-averaged WKB method and cross-examined via time-domain analyses within an acceptable parameter space. The results show that the estimated QNMs are more sensitive to $\ell$; however, both model parameters influence the frequency spectra.
The derived bounds on the GBFs aid in further constraining the parameter space. It is shown that $\ell$ and $\eta$ have a similar effect on the greybody bounds. Furthermore, positive and negative values of $\ell$ have opposing effects in that the bounds are reversed for the two cases. The analyses of the Hawking radiation sparsity highlight the effect of $\ell$, and two scenarios are noted: either the radiation emitted is less sparse than Hawking radiation, or it is more sparse during the evaporation phase. Thus, this work presents a comprehensive account of BHs in KR gravity with a global monopole charge.}

\begin{document}
\maketitle
\flushbottom

\section{Introduction}
\label{sec:int}
General Relativity (GR) is a well-tested theory that has withstood over a century of tests, ranging from the perihelion precession of Mercury to the detection of gravitational waves (GWs) and imaging of black hole (BH) shadows. However, GR has inherent limitations in that it cannot account for the dark sector of the universe and is a completely classical theory. Various modifications and/or extensions to GR have been proposed over the decades to address these limitations, and recent progress in observational astrophysics has opened the potential to explore deviations from GR. To this end, BHs serve as an intriguing testing ground. Stemming from the first reported solution to the field equations of GR in 1916, BHs are extremely compact objects formed during stellar collapse, at the cores of galaxies, and possibly due to density fluctuations in the early universe. Although state-of-the-art observations of BHs, such as those by the LIGO-Virgo-KAGRA and EHT collaborations, agree with GR \cite{LIGOScientific:2016aoc,LIGO16e,Event2021a,Event2021b}, strong motivations exist that hint at the possible detection of deviations at higher sensitivities. It is anticipated that such observations may be realized by planned missions such as the Cosmic Explorer, Einstein Telescope, Lisa Telescope, Tian Chen, Taiji, etc. In addition, studies on GWs and optical properties of BHs in modifications of GR is a subject of considerable interest. %Relevant citations to be included please%

Symmetries are of fundamental importance to physics, and Lorentz symmetry ensures that physical laws remain consistent across inertial frames of reference. However, theoretical evidence suggests that Lorentz symmetry may be violated at extremely high energies and/or small length scales \cite{collins2004lorentz, kostelecky1989spontaneous, alfaro2002loop, hovrava2009quantum, carroll2001noncommutative, jacobson2001gravity, dubovsky2005massive, bengochea2009dark, cohen2006very}. Such models of gravity contain variables/parameters that lead to Lorentz symmetry breaking (LSB) either explicitly or spontaneously. One the one hand, explicit violations are observed when the effective action contains functions of fixed vector or tensor fields that define a preferred direction or frame (the Lagrangian breaks Lorentz symmetry). On the other hand, spontaneous violations are observed when the vacuum expectation value (VEV) of a dynamical field breaks Lorentz symmetry. Thus, it is important to constrain such model parameters to ensure precision of future observations that may hint at the exact $4D$ theory of gravity. To this end, the Standard Model Extension (SME) framework is a well-studied paradigm for spontaneous LSB; SME includes a vector field called the bumblebee field \cite{bumblebee} with a non-zero VEV. BH and wormhole solutions in bumblebee gravity have been studied extensively in literature \cite{Maluf:2020kgf, Capozziello:2023rfv,Khodadi:2023yiw,Xu:2023xqh,Xu:2022frb,Oliveira:2021abg,Poulis:2021nqh,Casana:2017jkc,gullu2022schwarzschild, casana2018exact, Ovgun:2018xys,gogoi2022quasinormal}. Here, we focus instead on a higher-rank field of interest.

In addition to the bumblebee field, a rank-two antisymmetric field, called the Kalb-Ramond (KR) field, has been studied extensively \cite{KR,KRVEVAction}. Stemming from bosonic string theory, the non-minimal coupling of the KR field with gravity endows it with a non-zero VEV, leading to spontaneous, local LSB. The KR field may be interpreted as a generalization of the electromagnetic potential with two indices. It is well known that $p$-form theories \cite{bandos2021p} are generalizations of Maxwell's electrodynamics, with the gauge potential being a higher-rank differential form \cite{wald2010general}. In the KR model, the 2-form potential and field strength are given as $B_2=\frac{1}{2} B_{\mu \nu} dx^{\mu} \land dx^\nu$ and $H_{\alpha\mu\nu}=\partial_{[\alpha}B_{\mu\nu]}$, respectively. The KR field has been studied in the context of BH physics \cite{LorInvKRBH, sengupta2001spherically, kar2003static}, braneworld scenarios \cite{chakraborty2017strong}, cosmology \cite{krcmb, krgalactic}, etc. Static, spherically-symmetric BH in KR gravity have been reported \cite{lessa2020modified}, and relevant properties such as quasinormal modes (QNMs), light deflection, and particle dynamics have been studied \cite{baruah2023quasinormal, atamurotov2022particle,krlepto,kroptical}. %%Citations to be included for BB and KR literature%%.
An intriguing new solution in KR gravity has been reported very recently by Fathi and Ovg\"un \cite{fathi2025black}, where they consider the interaction of the KR field with a global monopole charge. Global monopoles are topological defects arising from the spontaneous breaking of global symmetries, specifically, $O(3) \rightarrow U(1)$ symmetry breaking, in the context of early-universe phase transitions \cite{vilenkin1981gravitational, vilenkin1982cosmological,dadhich1998schwarzschild}. They have significant implications both for cosmology and BH astrophysics. Global monopoles are generally associated with physical singularities; however, their pathological behaviors can be ameliorated when gravitational effects are taken into consideration \cite{barriola1989gravitational}. Especially, the self-energy divergence issue is resolved, highlighting its potential physical significance in astrophysical contexts \cite{bronnikov2002global}.  

In this study, we study some properties of the BH with a global monopole charge in KR gravity. Specifically, we study the QNMs, which are characteristic responses of BHs to perturbations using a high-accuracy method. Next, we derive strict bounds on the greybody factors (GBFs), which are crucial thermodynamic parameters concerning the radiative spectra of BHs. Finally, we examine the sparsity of Hawking radiation emitted from the BH. %%%More details may be added here.%%%

The remainder of this paper is organized as follows. In Section \ref{sec:krbh}, we review KR gravity in the context of its interaction with a global monopole and present the BH solution of interest. In Section \ref{sec:qnm}, we study the QNMs of the BH. In Section \ref{sec:gbfs}, we derive rigorous bounds on the GBFs. In Section \ref{sec:spar}, we discuss the sparsity of Hawking radiation of the KR BH, and finally, in Section \ref{sec:discn}, we present some discussions and conclude the paper.

Throughout the work, we paper in the natural system of units ($\kappa=8\pi G=c=\hbar = 1$) and the ($-,+,+,+$) sign convention.

\section{Black Hole in KR Gravity with a Global Monopole Chrage}
\label{sec:krbh}
The Einstein-Hilbert action minimally coupled with the self-interacting KR field is given by \cite{KRVEVAction, lessa2020modified}
\begin{eqnarray}
\label{eq:kractn}  
S=\frac{1}{2}\int d^4x \sqrt{-g} \left[R-\frac{1}{6}H^{\mu\nu\rho}H_{\mu\nu\rho}-V(B_{\mu\nu}B^{\mu\nu}) +  \xi_2 B^{\rho\mu}B^{\nu}_{\mu}R_{\rho\nu}+\xi_3 B^{\mu\nu}B_{\mu\nu}R\right] \\ \nonumber + \int d^4x \sqrt{-g} \mathcal{L}_M,
\end{eqnarray}
where $\xi_i$ represent the coupling constants. The field of interest here is the antisymmetric 2-tensor (KR field) $B_{\mu \nu} = - B_{\nu \mu}$, with its dual given by $\mathfrak{B}_{\mu \nu} \equiv \frac{1}{2} \epsilon_{\mu \nu \kappa \lambda} B^{\kappa \lambda}$, where $\epsilon_{\mu \nu \kappa \lambda}$ is the completely antisymmetric Levi--Civita tensor. Next, the field strength is given by the completely antisymmetric field--strength tensor defined in terms of derivatives of $B_{\mu \nu}$ as $H_{\lambda \mu \nu} = \partial_\lambda B_{\mu \nu} + \partial_\mu B_{\nu \lambda} + \partial_\nu B_{\lambda \mu}$. This can be interpreted as components of an exact 3-form field $H$ (the pseudoscalar axion) constructed from exterior derivatives of the 2-form $B$ \cite{KR, KRVEVAction}. Moreover, $\mathcal{L}_m$ denotes the matter Lagrangian corresponding to a global monopole, defined as \cite{barriola1989gravitational}:

\begin{equation}
    \mathcal{L}_m = \frac{1}{2} \partial_{\mu} \varphi^a \partial^{\mu} \varphi^a - \frac{\lambda}{4} \left(\varphi^a \varphi^a - \eta^2 \right)^2
\end{equation}
Here, $\lambda$ is a coupling constant and $\eta$ is the monopole charge. Equation \eqref{eq:kractn} can be varied with respect to $g^{\mu \nu}$ to obtain the gravitational field equation as:

\begin{equation}
    R_{\mu \nu} - \frac{1}{2} g_{\mu \nu} R = T^M_{\mu \nu} + T^{KR}_{\mu \nu}
\end{equation}

Considering a non-vanishing VEV for the KR field ($<B_{\mu \nu}> = b_{\mu \nu}$), the potential should be of the form $V = V(B^{\mu \nu}B_{\mu \nu} \pm b^2)$, using which the field equations becomes:

\begin{eqnarray}
    R_{\mu \nu} = T^M_{\mu \nu} - \frac{1}{2} g_{\mu \nu} T^M + V'(Y) + \xi_2 \left[g_{\mu \nu} b^{\alpha \gamma} b^\beta_\gamma R_{\alpha \beta} - b^{\alpha \beta} b_{\mu \beta} R_{\nu \alpha} - b^{\alpha \beta} b_{\nu \beta} R_{\mu \alpha} \right. \\ \left. \nonumber  + \frac{1}{2} \nabla_{\alpha} \nabla_{\mu} (b^{\alpha \beta} b_{{\nu \beta}}) + \frac{1}{2} \nabla_{\alpha} \nabla_{\nu} (b^{\alpha \beta} b_{{\mu \beta}}) - \frac{1}{2} \Box (b_{\mu}\,^{\gamma} b_{\nu \gamma})\right]
\end{eqnarray}
with $T^M = g^{\mu \nu} T^M_{\mu \nu}$ and $Y = 2 b_{\mu \alpha} b_{\nu}\,^{\alpha} + b^2 g_{\mu \nu}$. Moreover, the energy-momentum tensor corresponding to $\mathcal{L}_M$ is given by
\begin{eqnarray}
    T^M_{\mu \nu} = \partial_{\mu} \varphi^a \partial_{\nu} \varphi^a - g_{\mu \nu} \left[\frac{1}{2} \partial^{\rho} \varphi^a \partial_{\rho} \varphi^a - \frac{\lambda}{4} \left(\varphi^a \varphi^a - \eta^2 \right)^2 \right].
\end{eqnarray}

Using the static, spherically-symmetric metric ansatz
\begin{equation}
    ds^2 = -A(r)dt^2+B(r)dr^2+r^2d\theta^2+r^2\sin^2\theta d\phi^2,
\end{equation}

the following field equations are obtained
\begin{eqnarray}
    \frac{2 A'' (r)}{A(r)}  - \frac{A'(r)}{A(r)} \frac{B'(r)}{B(r)} -  \frac{A'(r)^2}{A(r)^2} + \frac{4}{r} \frac{A'(r)}{A(r)} = 0 \\
    \frac{2 A''(r)}{A(r)} - \frac{A'(r)}{A(r)} \frac{B'(r)}{B(r)} - \frac{A'(r)^2}{A(r)^2} - \frac{4}{r} \frac{B'(r)}{B(r)} = 0 \\ 
    \frac{2 A'' (r)}{A(r)}  - \frac{A'(r)}{A(r)} \frac{B'(r)}{B(r)} -  \frac{A'(r)^2}{A(r)^2} + \frac{1+\ell}{\ell r} \left[\frac{A'(r)}{A(r)} - \frac{B'(r)}{B(r)} \right] \\ \nonumber - \left[1 - b^2 r^2 V'(Y) \right] \frac{2 B(r)}{\ell r^2} + \frac{2(1-\ell}{\ell r ^2} + \eta^2 = 0
\end{eqnarray}
with $\ell \equiv \xi_2 b^2 /2$. Considering the VEV at the local minimum of the potential \cite{yang2023static} and substracting Eq. (8) from (7) leads to $A(r) = B(r)^{-1}$. Finally, substracting Eq. (9) from (7) provides the solution for $A(r)$, which in the second order of the monopole charge is 
\begin{equation}
    A(r) = \frac{1}{1-\ell} - \frac{2M}{r} + \frac{\ell M \eta^2 r}{2(1-\ell)} - \frac{\ell \eta^2 r^2}{6 (1-\ell)^2} + \mathcal{O} (\eta^4).
\end{equation}

This BH with a global monopole charge has been studied in terms of the horizon properties, thermodynamics, and solar system tests, constraining the acceptable ranges of $\ell$ and $\eta$ \cite{fathi2025black}. In the next sections, we study the QNMs, GBFs, and sparsity of Hawking radiation for this BH.

\section{Quasinormal Modes}
\label{sec:qnm}
In this section, we study the QNMs of the KR BH with a global monopole charge. QNMs quantify the responses of BHs to perturbations of test fields. It should be noted that the back-reaction of the fields on the space-time is considered negligible in such analyses. We first study the variations of the QNMs corresponding to scalar, electromagnetic (EM), and gravitational perturbations of the KR BH with the model parameters in the frequency domain using the 13\textsuperscript{th}-order Pad\'{e}-averaged WKB method. The, we study the time-domain profiles of the perturbations and extract the dominant QNMs for each case. We start with a brief introduction of the Pad\'{e}-averaged WKB method.

\subsection{The Pad\'{e}-averaged WKB method for QNMs}
\label{sec:padeqnm}
The WKB method a well-established approach to solve BH perturbation equations in the frequency domain \cite{mashoon, schutz1985black, iyer1987black, Konoplya6thOrder}. However, the accuracy of the estimated frequencies degrade for $n \geq l$. To ameliorate this issue, the Pad\'{e} approximation \cite{matyjasekopalaWKB} can be used to evaluate QNMs with higher precision. Here, we use the Pad\'{e}-averaged WKB approach, as briefly outlined below, to estimate the frequency-domain QNMs for the KR BH subjected to scalar, EM, and gravitational perturbations. 

For a wave-like equation
\begin{equation}
    \label{eq:wave}
    \frac{d^2 \Psi}{dx^2} = U (x, \omega) \Psi,
\end{equation}
the WKB method yields solutions in asymptotic regions described by a superposition of ingoing and outgoing waves \cite{konoplya2011quasinormal}. The asymptotic solutions are matched with the extrema of the effective potential via a Taylor expansion. The WKB method provides a closed--form formula for the QNM frequencies as \cite{konoplya2019higher}
\begin{equation}
    \label{eq:bhwkb}
    \omega^2=V_0+A_2(\mathcal{K}^2)+A_4(\mathcal{K}^2)+A_6(\mathcal{K}^2)+\ldots- i  \mathcal{K}\sqrt{-2V_2}\left(1+A_3(\mathcal{K}^2)+A_5(\mathcal{K}^2)+A_7(\mathcal{K}^2)\ldots\right),
\end{equation}
% where $\mathcal{K}$ admits half--integer values
% \begin{eqnarray}
% \mathcal{K} &=& \left\{
% \begin{array}{ll}
%  +n+\frac{1}{2}, & Re(\omega)>0; \\
%  -n-\frac{1}{2}, & Re(\omega)<0; \phantom{\frac{{}^{Whitespace}}{}}
% \end{array}
% \right.\\\nonumber
% &&\qquad\quad\qquad n=0,1,2,3\ldots.
% \end{eqnarray}
The divergence of the Taylor series is avoiuded by using Pad\'{e} approximants \cite{matyjasekopalaWKB}, where a polynomial $P_k(\epsilon)$, defined in powers of an \emph{order parameter} $\epsilon$, modifies Eq. \eqref{eq:bhwkb} as
\begin{equation}
P_k(\epsilon)=V_0+A_2(\mathcal{K}^2)\epsilon^2+A_4(\mathcal{K}^2)\epsilon^4+A_6(\mathcal{K}^2)\epsilon^6+\ldots - i \mathcal{K}\sqrt{-2V_2}\left(\epsilon+A_3(\mathcal{K}^2)\epsilon^3+A_5(\mathcal{K}^2)\epsilon^5\ldots\right)
\end{equation}
the Pad\'{e} approximants $P_{\tilde{n}/\tilde{m}}(\epsilon)$ for the polynomial $P_k(\epsilon)$ is given by rational functions \cite{matyjasekopalaWKB, konoplya2019higher}:
\begin{equation}
    P_{\tilde{n}/\tilde{m}}(\epsilon)=\frac{Q_0+Q_1\epsilon+\ldots+Q_{\tilde{n}}\epsilon^{\tilde{n}}}{R_0+R_1\epsilon+\ldots+R_{\tilde{m}}\epsilon^{\tilde{m}}},
\end{equation}
with $\tilde{n}+\tilde{m}=k$. To estimate the accuracy of the approach, we calculate the associated errors in the frequencies. In the WKB formula used, corrections in each order affect the real and imaginary parts of the squared frequency, and the error in $\omega_k$ for an arbitrary order $k$ can be estimated as
\begin{equation}
    \Delta_k = \frac{|\omega_{k+1} - \omega_{k-1}|}{2}
\end{equation}

\subsection{Massless Scalar Perturbations}
Considering massless scalar perturbations, we start with the Klein--Gordon equation; the perturbed metric can be recast without loss of generality as follows \cite{bouhmadi2020consistent}:
\begin{equation}
    ds^2 = -|g_{tt}| dt^2 + g_{rr}dr^2 + r^2 d\theta^2 + r^2\sin^2 \theta \left(d\phi - a dt - bdr - c d\theta \right)^2
    \label{pertmetric}
\end{equation}

Here, $a$, $b$, and $c$ are functions of $t$, $r$, and $\theta$, encoding the perturbations. We leverage the tetrad formalism and adopt a basis $e^\mu_{a}$ associated with the metric $g_{\mu\nu}$, and satisfying
\begin{align}
e^{(a)}_\mu e^\mu_{(b)} &= \delta^{(a)}_{(b)} \notag \\
e^{(a)}_\mu e^\nu_{(a)} &= \delta^{\nu}_{\mu} \notag \\
e^{(a)}_\mu &= g_{\mu\nu} \eta^{(a)(b)} e^\nu_{(b)}\notag \\
g_{\mu\nu} &= \eta_{(a)(b)}e^{(a)}_\mu e^{(b)}_\nu = e_{(a)\mu} e^{(a)}_\nu.
\end{align}

In the new basis, vector and tensor quantities are projected as
\begin{align}
P_\mu &= e^{(a)}_\mu P_{(a)}, \notag\\ 
P_{(a)} &= e^\mu_{(a)} P_\mu, \notag\\
A_{\mu\nu} &=  e^{(a)}_\mu e^{(b)}_\nu A_{(a)(b)}, \notag\\
A_{(a)(b)} &= e^\mu_{(a)} e^\nu_{(b)} A_{\mu\nu}.
\end{align}
Considering the propagation of a massless scalar field around the black hole and assuming that the reaction of 
the scalar field on the space-time is negligible, the scalar QNMs can be described by the Klein--Gordon equation given by
\begin{equation}\label{scalar_KG}
\square \Phi = \dfrac{1}{\sqrt{-g}} \partial_\mu (\sqrt{-g} g^{\mu\nu} \partial_\nu \Phi) = 0.
\end{equation}

Neglecting the back--reaction of the field, one can consider Eq. \eqref{pertmetric} only up to the zeroth order:
\begin{equation}
    ds^2 = -|g_{tt}| dt^2 + g_{rr}dr^2 + r^2 d\Omega_2^2
\end{equation}

The scalar field can conventionally be decomposed using spherical harmonics as
\begin{equation}
\Phi(t,r,\theta, \phi) = \dfrac{1}{r} \sum_{l,m} \psi_l(t,r) Y_{lm}(\theta, \phi),
\end{equation}

where $\psi_l(t,r)$ is the time--dependent radial wave function. Further, $l$ and $m$ are 
the indices of the spherical harmonics $Y_{lm}$. Then, Eq. \eqref{scalar_KG} yields
\begin{equation}
\partial^2_{r_*} \psi(r_*)_l + \omega^2 \psi(r_*)_l = V_s(r) \psi(r_*)_l,  
\end{equation}

where $r_*$ is the tortoise coordinate defined as
\begin{equation}\label{tortoise}
\dfrac{dr_*}{dr} = \sqrt{g_{rr}\, |g_{tt}^{-1}|}
\end{equation}

and $V_s(r)$ is the effective potential of the field given by
\begin{equation}\label{Vs}
V_s(r) = |g_{tt}| \left( \dfrac{l(l+1)}{r^2} +\dfrac{1}{r \sqrt{|g_{tt}| g_{rr}}} \dfrac{d}{dr}\sqrt{|g_{tt}| g_{rr}^{-1}} \right).
\end{equation}

\begin{figure}[htbp]
     \centering
     \begin{subfigure}[t]{0.49\textwidth}
         \centering
         \includegraphics[width=\textwidth]{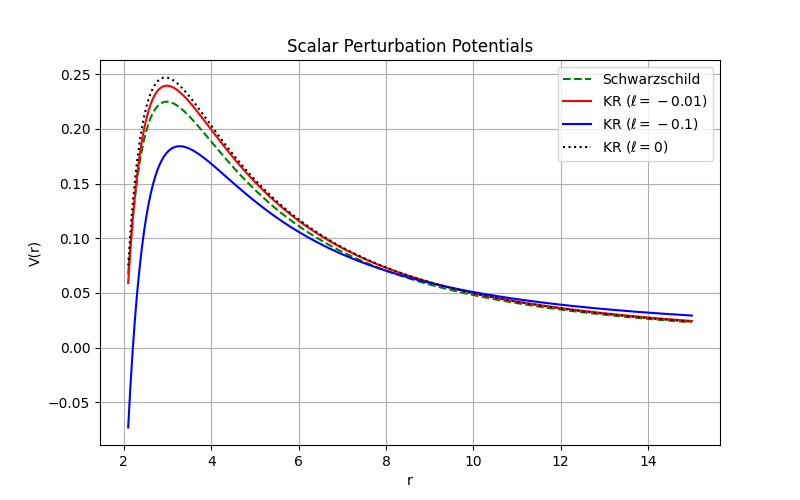}
         \caption{Variation with $\ell$}
         \label{fig:potvarl}
     \end{subfigure}
     %\hfill
     \begin{subfigure}[t]{0.49\textwidth}
         \centering
         \includegraphics[width=\textwidth]{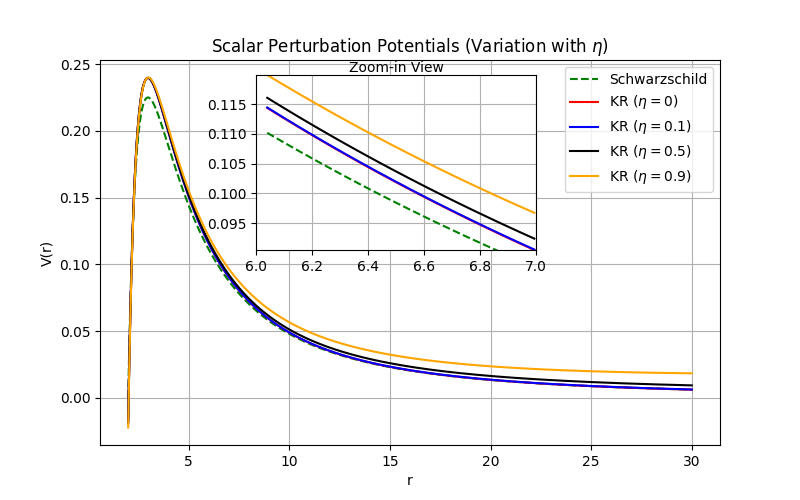}
         \caption{Variation with $\eta$}
         \label{fig:potvareta}
     \end{subfigure}
    \caption{Variation of the scalar potential with $\ell$ and $\eta$.}
    \label{fig:potvar}
\end{figure}

The effect of $\ell$ and $\eta$ on the scalar potential has been visualized in Fig. \ref{fig:potvar}.
\subsubsection{Variation of scalar QNMs with $\ell$}
\begin{figure}[htbp]
     \centering
     \begin{subfigure}[b]{\textwidth}
         \centering
         \includegraphics[width=\textwidth]{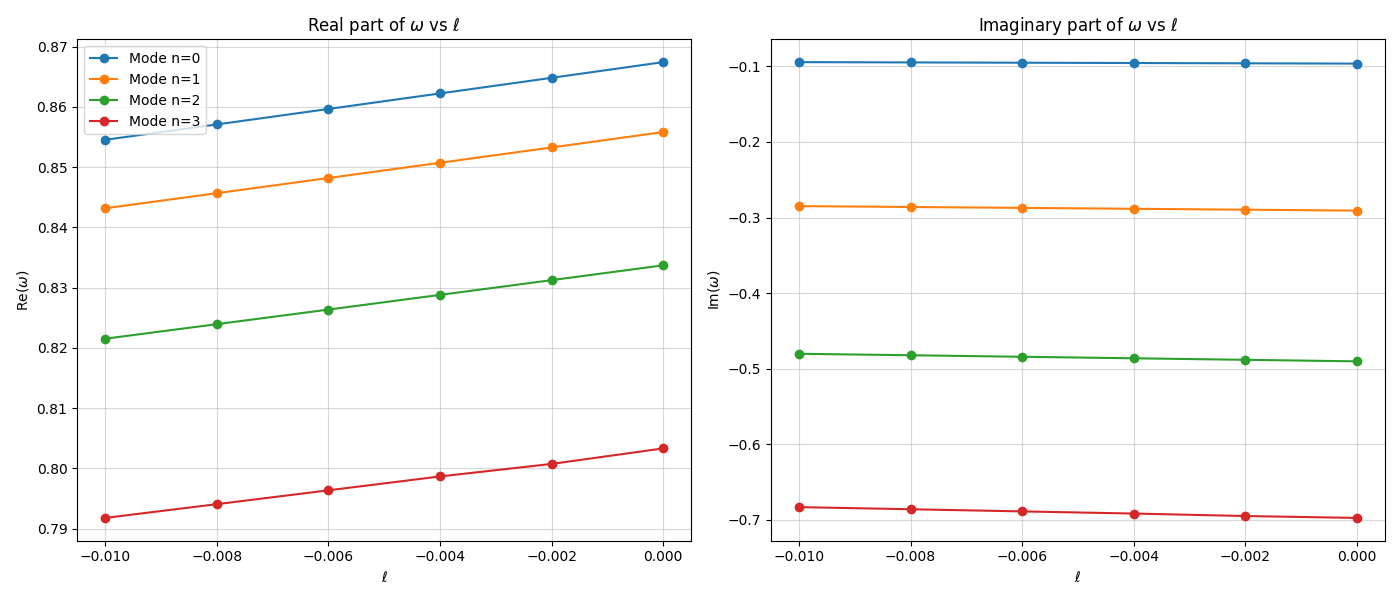}
         %\caption{Variation with $\ell$}
         %\label{fig:potvarl}
     \end{subfigure}
     %\hfill
     \begin{subfigure}[b]{\textwidth}
         \centering
         \includegraphics[width=\textwidth]{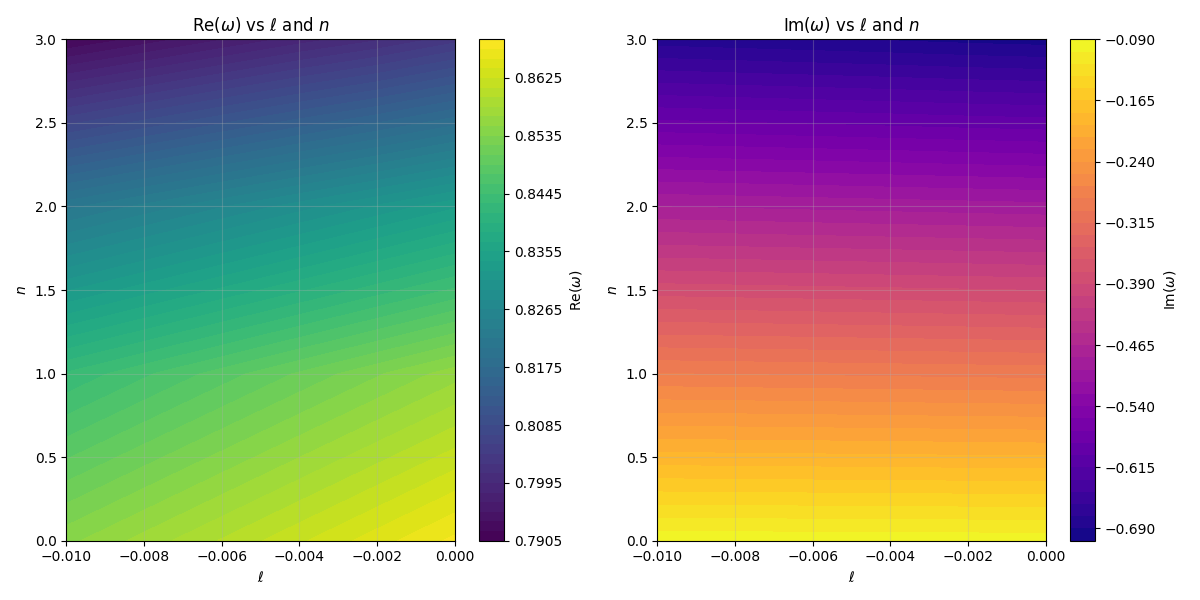}
         %\caption{Variation with $\eta$}
         %\label{fig:potvareta}
     \end{subfigure}
    \caption{Variation of the $l=4$ scalar QNMs with $\ell$ and fixed $\eta = 0.3$.}
    \label{fig:qnmvarellscalar}
\end{figure}

\begin{figure}
    \centering
    \includegraphics[width=0.7\textwidth]{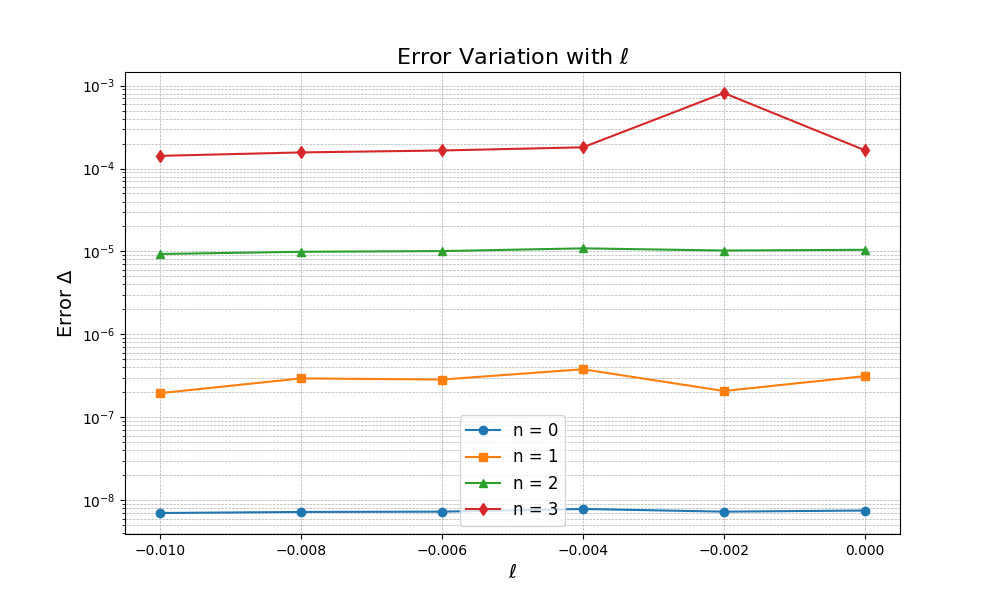}
    \caption{Errors in Table \ref{tab:qnmvarl}}
    \label{fig:errorellscalar}
\end{figure}

The variation of the scalar QNMs with $\ell$ is evident from the data listed in Table \ref{tab:qnmvarl} and visualized in Fig. \ref{fig:qnmvarellscalar}. It can be seen that $\omega_{\text{Re}}$ exhibits a monotonic increase as $\ell$ is varied from $-0.01$ to $0$. For example, for $n=0$ (the fundamental mode), $\omega_{\text{Re}}$ increases from approximately $0.8545$ to $0.8674$. Similar increasing trends are seen for higher overtones. The systematic trend in the real part indicates that the oscillation frequency of the QNMs is sensitive to the parameter $\ell$ in that as $\ell$ increases (or in our parameterization, becomes less negative), the oscillatory response becomes slightly more rapid. The imaginary part, which is associated with the damping rate of the mode, also exhibits systematic variation. For instance, for the fundamental mode, $\omega_{\text{Im}}$ becomes more negative, from approximately $-0.09442$ to $-0.09639$ with increasing $\ell$. Similar trends are observed for the higher overtones, suggesting that the damping becomes slightly more efficient as $\ell$  increases. Physically, this indicates that the QNMs decay faster as $\ell \rightarrow 0$. Across all values of $\ell$, the modes corresponding to higher overtone numbers exhibit both lower $\omega_{\text{Re}}$ and more negative $\omega_{\text{Im}}$ compared to those with lower $n$. This hierarchical structure is typical in QNM spectra and confirms that the influence of $\ell$ is modulated by the overtone index. Thus, the oscillatory frequency and damping rate exhibit systematic variations with $n$. The relative error $\Delta$ associated with each frequency has also been computed and visualized in Fig. \ref{fig:errorellscalar}; its logarithmic scaling with $\ell$ suggests that the numerical uncertainties are small but vary with with $\ell$. The error estimates confirm that the trends observed in the QNMs are robust over the range of $\ell$ values considered. The systematic behavior is not an artifact of numerical noise, but reflects actual physical dependence on $\ell$. Overall, it can be seen that the 13\textsuperscript{th}-order Pad\'{e}-averaged WKB method yields accurate results.
\begin{table}[htbp]
\centering
\begin{tabular}{|c|c|c|c|}
    \hline
    $\ell$ & $n$ & $\omega$  & $\Delta$ \\
    \hline
    -0.01 & 0 & $0.854519\, -0.0944194 i$ & $7.03513\times10^{-9}$ \\
   & 1 & $0.843157\, -0.284922 i$ & $1.96032\times10^{-7}$ \\
    & 2 & $0.82151\, -0.480279 i$ & $9.31994\times10^{-6}$ \\
    & 3 & $0.791767\, -0.683183 i$ & $0.000142056$ \\
    \hline
   -0.008 & 0 & $0.857073\, -0.0948092 i$ & $7.2419\times10^{-9}$ \\
   & 1 & $0.845662\, -0.286099 i$ & $2.9498\times10^{-7}$ \\
   & 2 & $0.823922\, -0.482263 i$ & $9.9095\times10{-6}$ \\
   & 3 & $0.79405\, -0.686006 i$ & 0.000156267 \\
    \hline    
    -0.006 & 0 & $0.859639\, -0.0952013 i$ & $7.30235\times10^{-9}$ \\
    & 1 & $0.848179\, -0.287282 i$ & $2.85788\times10^{-7}$ \\
    & 2 & $0.826347\, -0.484261 i$ & 0.000010104 \\
    & 3 & $0.796348\, -0.688848 i$ & 0.000164913 \\
    \hline
     -0.004 & 0 & $0.862218\, -0.0955957 i$ & $7.86920\times10^{-9}$ \\
    & 1 & $0.85071\, -0.288473 i$ & $3.80260\times10^{-7}$ \\
    & 2 & $0.828782\, -0.486269 i$ & 0.0000109012 \\
    & 3 & $0.79865\, -0.691702 i$ & 0.000180047 \\  
    \hline
    -0.002 & 0 &$ 0.86481\, -0.0959925 i$ & $7.30356\times10^{-9}$ \\
    & 1 & $0.853252\, -0.289671 i$ & $2.07926\times10^{-7}$ \\
    & 2 & $0.831232\, -0.488294 i$ & 0.0000102581 \\
    & 3 & $0.800732\, -0.694983 i$ & 0.000813118 \\
    \hline
    0 & 0 & $0.867416\, -0.0963917 i$ & $7.53890\times10^{-9}$ \\
    & 1 & $0.855808\, -0.290876 i$ & $3.13969\times10^{-7}$ \\
    & 2 & $0.833692\, -0.490326 i$ & 0.0000104755 \\
    & 3 & $0.803307\, -0.6975 i$ & 0.000166087 \\
    \hline
\end{tabular}
\caption{Variation of $l=4$ scalar QNMs with $\ell$ and fixed $\eta =0.3$}
\label{tab:qnmvarl}
\end{table}
\subsubsection{Variation of scalar QNMs with $\eta$}
Next, we analyze the behavior of the scalar QNMs with $\eta$ while keeping $\ell$ fixed for four overtone numbers ($n = 0, 1, 2, 3$) over the range $\eta \in [0, 0.9]$. It can be seen that $\omega_{\mathrm{Re}}$ exhibits a slight, monotonic increase with increasing $\eta$ for each overtone. Although the changes are subtle, typically on the order of $10^{-4}$, the trend is consistent across all overtones and possibly, indicative of a weak dependence on $\eta$. In contrast, $\omega_{\mathrm{Im}}$, exhibits a systematic decrease in its absolute magnitude with increasing $\eta$. Physically, this reduction in $|\omega_{\mathrm{Im}}|$ implies that the damping reduces (or equivalently, decay time increases) with increasing $\eta$. This behavior may be indicative of a weakening of the dissipative effects in the system. Moreover, for the lower modes ($n=0$ and $n=1$), the errors remain consistently small (on the order of $10^{-9}$ and $10^{-7}$, respectively) across the studied range of $\eta$. However, for the higher overtones ($n=2$ and $n=3$), notable deviations at specific values of $\eta$ are observed. In particular, an anomalous increase in the error for the $n=2$ mode at $\eta \approx 0.5$ and for the $n=3$ mode near $\eta \approx 0.7$ suggests increased sensitivity of the WKB method.
\begin{table}[htbp]
\centering
\begin{tabular}{|c|c|c|c|}
    \hline
    $\eta$ & $n$ & $\omega$  & $\Delta$ \\
    \hline
    0 & 0 & $0.8545\, -0.0944908 i$ & $7.08669\times10^{-9}$ \\
      & 1 & $0.843174\, -0.285124 i$ & $2.91345\times10^{-7}$ \\
      & 2 & $0.821584\, -0.480579 i$ & $0.000010025$ \\
      & 3 & $0.791894\, -0.683541 i$ & $0.000160666$ \\
    \hline
    0.1 & 0 & $0.854502\, -0.0944829 i$ & $7.03077\times10^{-9}$ \\
        & 1 & $0.843172\, -0.285102 i$ & $2.061503\times10^{-7}$ \\
        & 2 & $0.821576\, -0.480547 i$ & $9.83541\times10^{-6}$ \\
        & 3 & $0.791882\, -0.683507$ i & $0.000154594$ \\
    \hline
    
    0.2 & 0 & $0.854508\, -0.0944591 i$ & $7.06444\times10^{-9}$ \\
        & 1 & $0.843166\, -0.285034 i$ & $2.71866\times10^{-7}$ \\
        & 2 & $0.821551\, -0.480445 i$ & $9.76061\times10^{-6}$ \\
        & 3 & $0.791836\, -0.68338 i$ & $0.000154112$ \\
    \hline
    0.3 & 0 & $0.854519\, -0.0944194 i$ & $7.03512\times10^{-9}$ \\
        & 1 & $0.843157\, -0.284922 i$ & $1.96032\times10^{-7}$ \\
        & 2 & $0.82151\, -0.480279 i$ & $9.31993\times10^{-6}$ \\
        & 3 & $0.791767\, -0.683183 i$ & $0.000142056$ \\
    \hline
    0.4 & 0 & $0.854535\, -0.0943638 i$ & $7.10626\times10^{-9}$ \\
        & 1 & $0.843144\, -0.284765 i$ & $1.47834\times10^{-7}$ \\
        & 2 & $0.821452\, -0.480043 i$ & $9.67111\times10^{-6}$ \\
        & 3 & $0.791665\, -0.682886 i$ & $0.00015724$ \\
    \hline
    0.5 & 0 & $0.854557\, -0.0942922 i$ & $7.78794\times10^{-9}$ \\
        & 1 & $0.843128\, -0.284563 i$ & $3.59920\times10^{-7}$ \\
        & 2 & $0.821696\, -0.47965 i$ & $0.000668629$ \\
        & 3 & $0.791528\, -0.682512 i$ & $0.000162645$ \\
    \hline    
    0.6 & 0 & $0.854585\, -0.0942048 i$ & $2.31978\times10^{-8}$ \\
        & 1 & $0.843111\, -0.284316 i$ & $7.11556\times10^{-8}$ \\
        & 2 & $0.82129\, -0.479375 i$ & $8.70463\times10^{-6}$ \\
        & 3 & $0.791376\, -0.68208 i$ & $0.000139143$ \\
    \hline
    0.7 & 0 & $0.854622\, -0.0941012 i$ & $7.07419\times10^{-9}$ \\
        & 1 & $0.843092\, -0.284024 i$ & $2.47424\times10^{-7}$ \\
        & 2 & $0.821185\, -0.478938 i$ & $9.307742\times10^{-6}$ \\
        & 3 & $0.790512\, -0.682186 i$ & $0.00176716$ \\
    \hline
    0.8 & 0 & $0.854666\, -0.0939816 i$ & $6.38668\times10^{-9}$ \\
        & 1 & $0.843073\, -0.283687 i$ & $8.82210\times10^{-8}$ \\
        & 2 & $0.821069\, -0.47844 i$ & $9.04664\times10^{-6}$ \\
        & 3 & $0.790982\, -0.680945 i$ & $0.000144749$ \\
    \hline
    0.9 & 0 & $0.854721\, -0.0938459 i$ & $7.16993\times10^{-9}$ \\
        & 1 & $0.843055\, -0.283305 i$ & $1.63755\times10^{-7}$ \\
        & 2 & $0.820935\, -0.477872 i$ & $8.39504\times10^{-6}$ \\
        & 3 & $0.790726\, -0.680256 i$ & $0.000134466$ \\
    \hline
\end{tabular}
\caption{Variation of $l=4$ scalar QNMs with $\eta$ and fixed $\ell =-0.01$}
\label{tab:qnmvareta}
\end{table}
\begin{figure}[htbp]
     \centering
     \begin{subfigure}[b]{\textwidth}
         \centering
         \includegraphics[width=\textwidth]{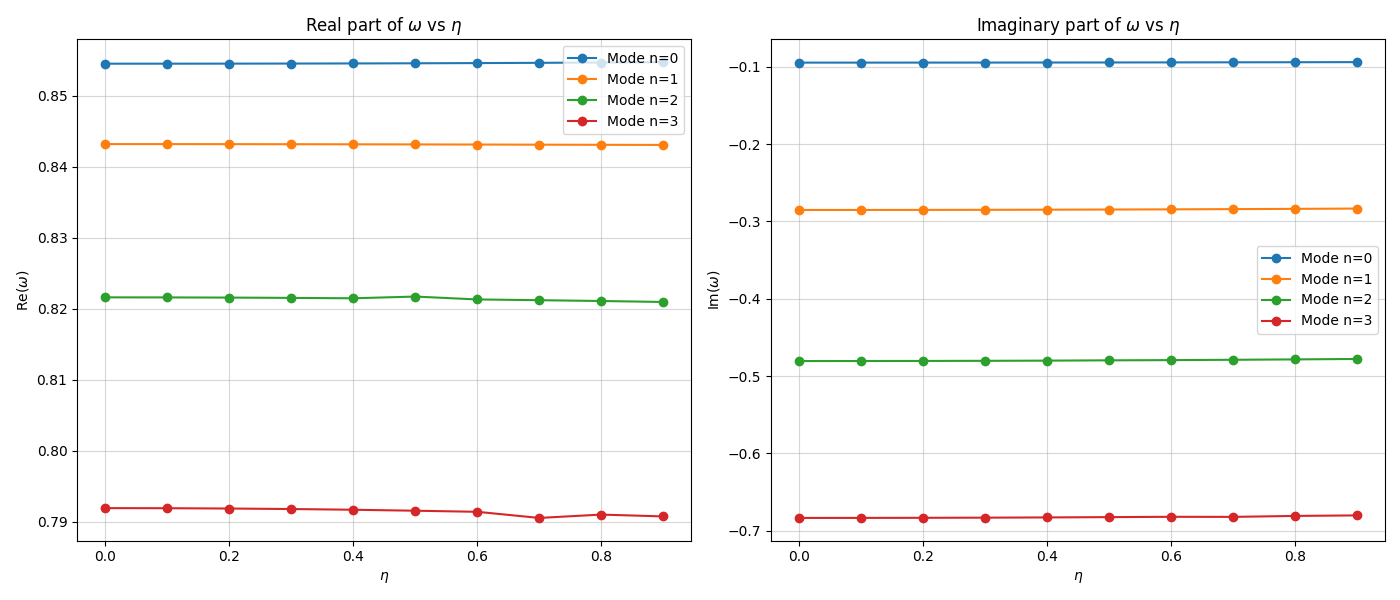}
         %\caption{Variation with $\ell$}
         %\label{fig:potvarl}
     \end{subfigure}
     %\hfill
     \begin{subfigure}[b]{\textwidth}
         \centering
         \includegraphics[width=\textwidth]{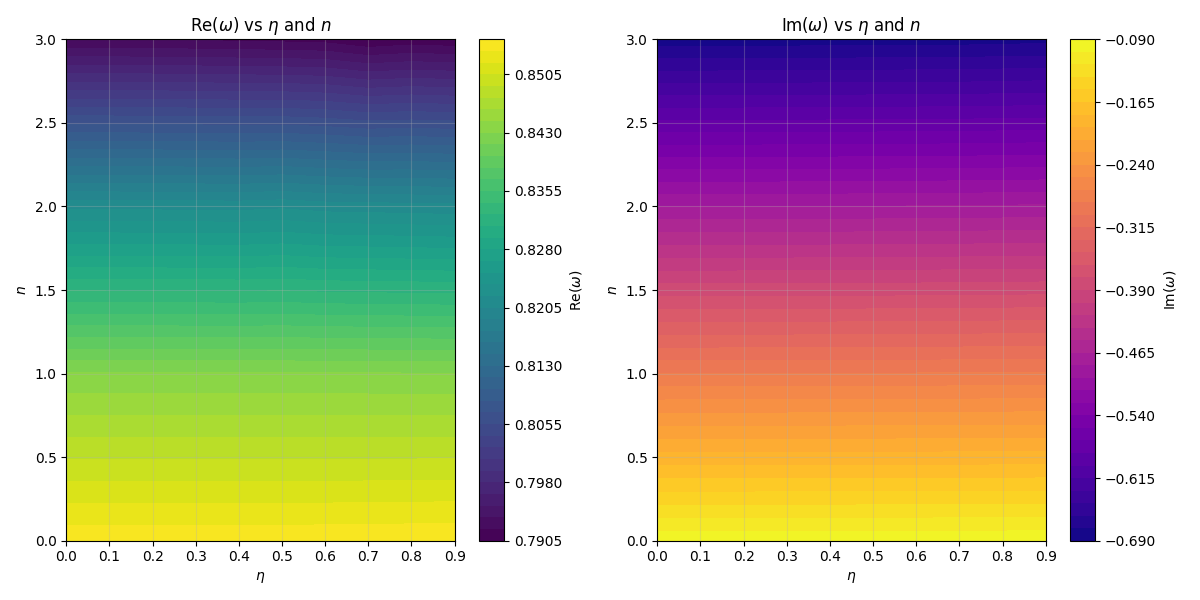}
         %\caption{Variation with $\eta$}
         %\label{fig:potvareta}
     \end{subfigure}
    \caption{Variation of the $l=4$ scalar QNMs with $\eta$ and fixed $\ell = -0.01$.}
    \label{fig:qnmvaretascalar}
\end{figure}
\begin{figure}
    \centering
    \includegraphics[width=0.7\textwidth]{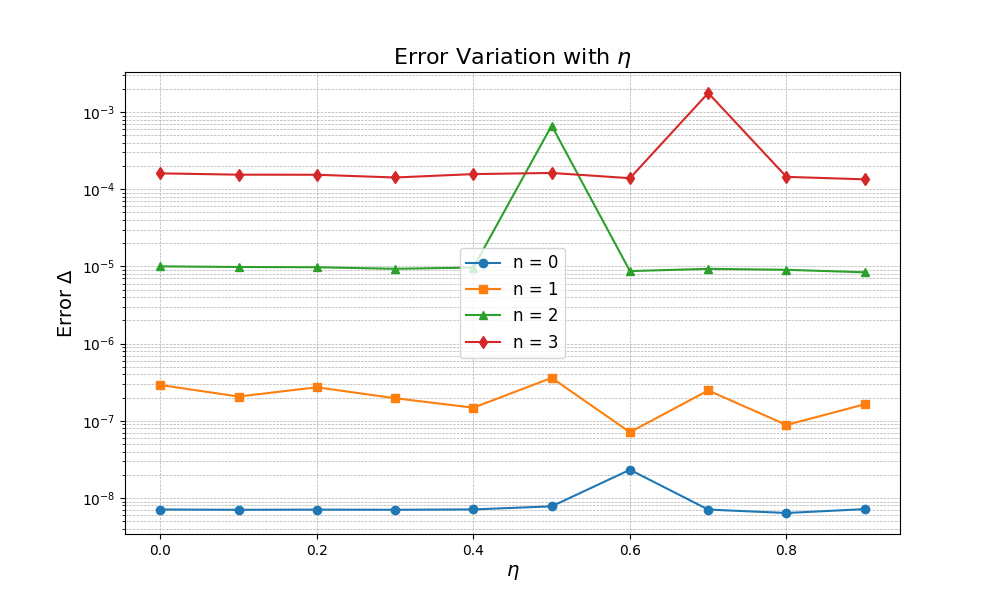}
    \caption{Errors in Table \ref{tab:qnmvareta}}
    \label{fig:erroretascalar}
\end{figure}
\subsection{Electromagnetic Perturbations}
Here, we consider electromagnetic (EM) perturbations on the KR BH in the tetrad formalism \cite{chandrasekhar1991selected}. Using the Bianchi identity of the EM field strength $F_{[(a)(b)(c)]} = 0$, we have:
\begin{align}
\left( r \sqrt{|g_{tt}|}\, F_{(t)(\phi)}\right)_{,r} + r \sqrt{g_{rr}}\, F_{(\phi)(r), t} &=0, \label{em1} \\
\left( r \sqrt{|g_{tt}|}\, F_{(t)(\phi)}\sin\theta\right)_{,\theta} + r^2 \sin\theta\, F_{(\phi)(r), t} &=0. \label{em2}
\end{align}
The conservation equation, $\eta^{(b)(c)}\! \left( F_{(a)(b)} \right)_{|(c)} =0$, gives
\begin{equation} \label{em3}
\left( r \sqrt{|g_{tt}|}\, F_{(\phi)(r)}\right)_{,r} +  \sqrt{|g_{tt}| g_{rr}}\, F_{(\phi)(\theta),\theta} + r \sqrt{g_{rr}}\, F_{(t)(\phi), t} = 0.
\end{equation}
Redefining the field perturbation as $\mathcal{F} = F_{(t)(\phi)} \sin\theta$, Eq. \eqref{em3} can be differentiated and used in Eqs. \eqref{em1} and \eqref{em2} to obtain
\begin{equation}\label{em4}
\left[ \sqrt{|g_{tt}| g_{rr}^{-1}} \left( r \sqrt{|g_{tt}|}\, \mathcal{F} \right)_{,r} \right]_{,r} + \dfrac{|g_{tt}| \sqrt{g_{rr}}}{r} \left( \dfrac{\mathcal{F}_{,\theta}}{\sin\theta} \right)_{,\theta}\!\! \sin\theta - r \sqrt{g_{rr}}\, \mathcal{F}_{,tt} = 0,
\end{equation}

Then, using the Fourier and field decompositions, $(\partial_t \rightarrow -\, i \omega)$ and $\mathcal{F}(r,\theta) = \mathcal{F}(r) Y_{,\theta}/\sin\theta$, respectively\footnote{Here, $Y(\theta)$ is the Gegenbauer function \cite{abramowitz1964handbook}.} \cite{chandrasekhar1991selected}, Eq. \eqref{em4} can be recast as
\begin{equation}\label{em5}
\left[ \sqrt{|g_{tt}| g_{rr}^{-1}} \left( r \sqrt{|g_{tt}|}\, \mathcal{F} \right)_{,r} \right]_{,r} + \omega^2 r \sqrt{g_{rr}}\, \mathcal{F} - |g_{tt}| \sqrt{g_{rr}} r^{-1} l(l+1)\, \mathcal{F} = 0.
\end{equation}
Finally, using $\psi_e \equiv r \sqrt{|g_{tt}|}\, \mathcal{F}$ and introducing the tortoise coordinate, the perturbation equation is written as a Schr\"odinger-like equation as follows
\begin{equation}
\partial^2_{r_*} \psi_e + \omega^2 \psi_e = V_e(r) \psi_e,
\end{equation}
with the potential defined as
\begin{equation}\label{Ve}
V_e(r) = |g_{tt}|\, \dfrac{l(l+1)}{r^2}. 
\end{equation}

\begin{figure}[htbp]
     \centering
     \begin{subfigure}[t]{0.49\textwidth}
         \centering
         \includegraphics[width=\textwidth]{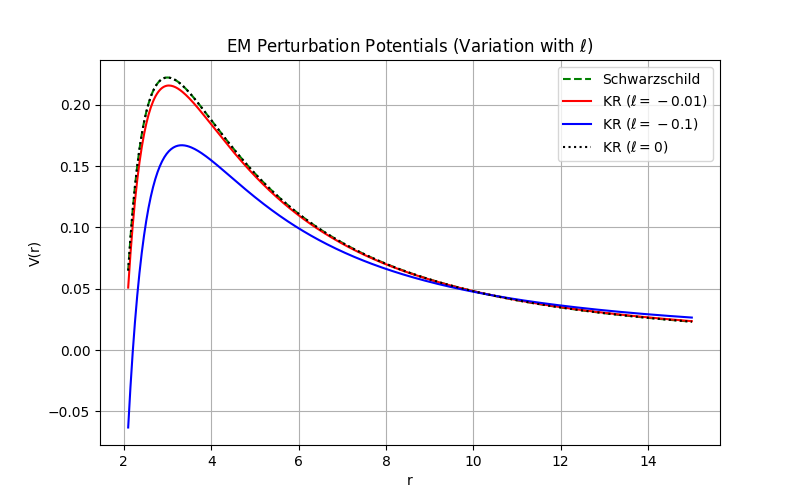}
         \caption{Variation with $\ell$}
         \label{fig:empotvarl}
     \end{subfigure}
     \begin{subfigure}[t]{0.49\textwidth}
         \centering
         \includegraphics[width=\textwidth]{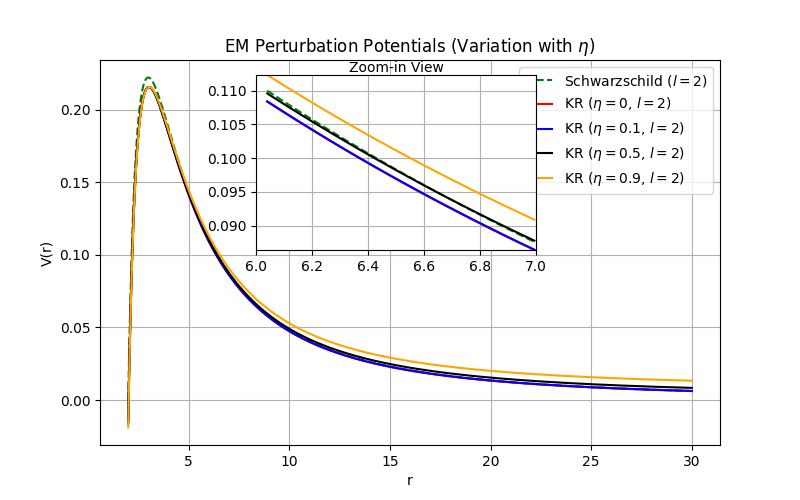}
         \caption{Variation with $\eta$}
         \label{fig:empotvareta}
     \end{subfigure}
    \caption{Variation of the EM potential with $\ell$ and $\eta$.}
    \label{fig:empotvar}
\end{figure}

The effect of $\ell$ and $\eta$ on the EM potential has been visualized in Fig. \ref{fig:empotvar}.

\subsubsection{Variation of EM QNMs with $\ell$}
\begin{figure}[htbp]
     \centering
     \includegraphics[width=\textwidth]{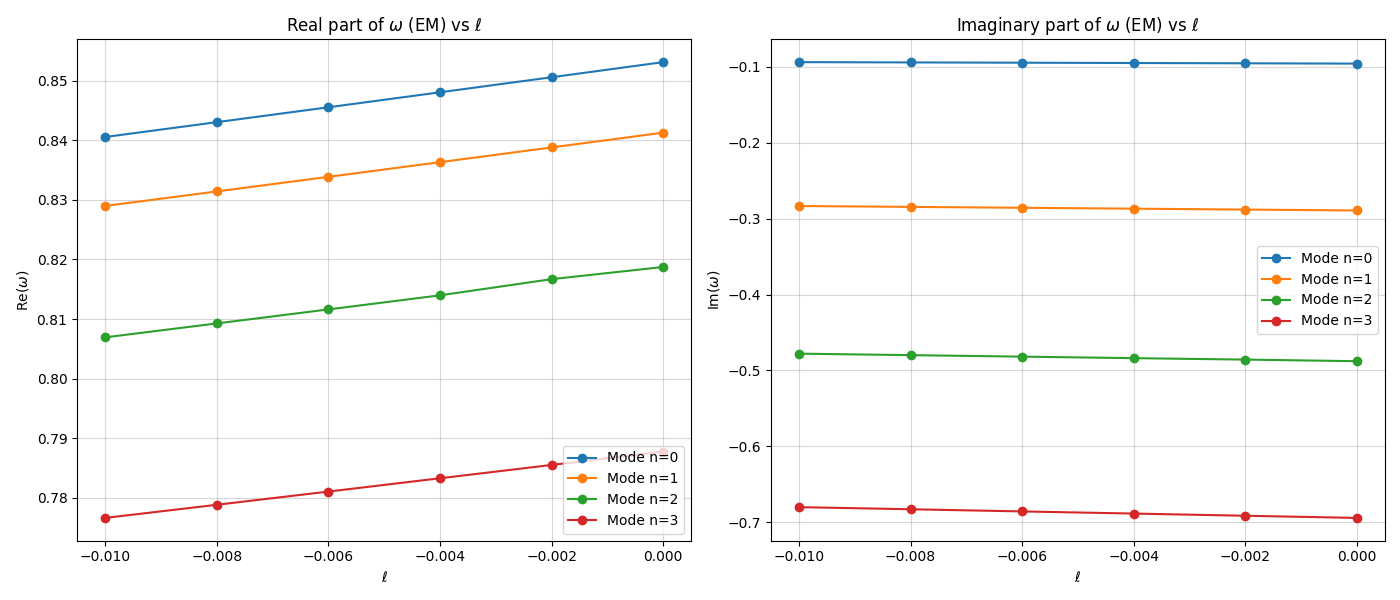}
     \includegraphics[width=\textwidth]{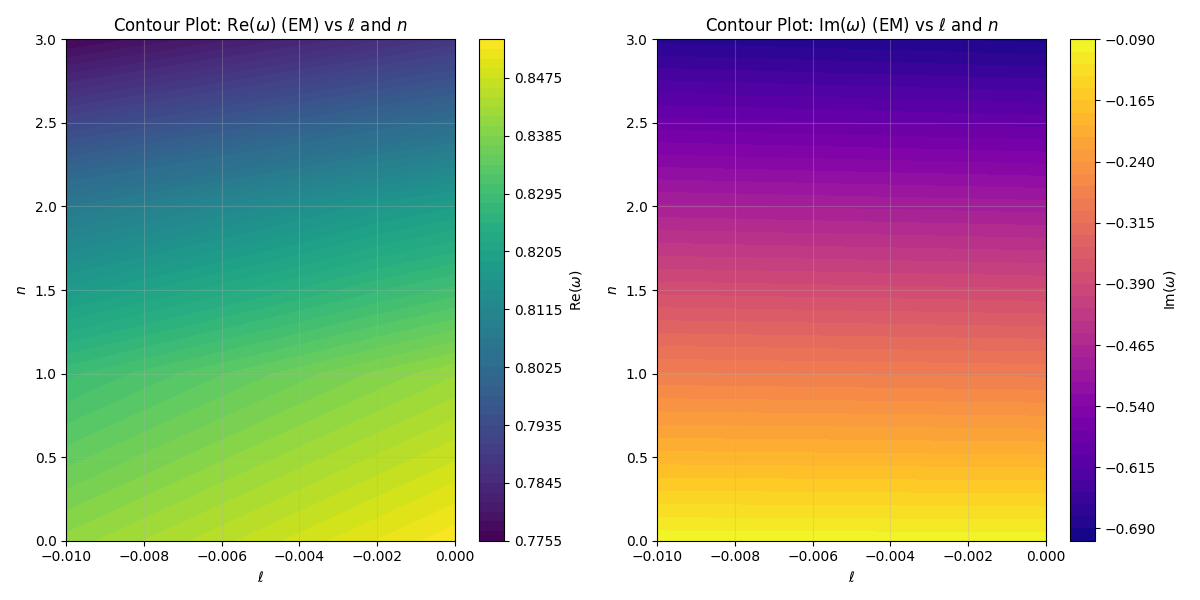}
    \caption{Variation of the $l=4$ EM QNMs with $\ell$ and fixed $\eta = 0.3$.}
    \label{fig:emqnmvarell}
\end{figure}

\begin{figure}
    \centering
    \includegraphics[width=0.7\textwidth]{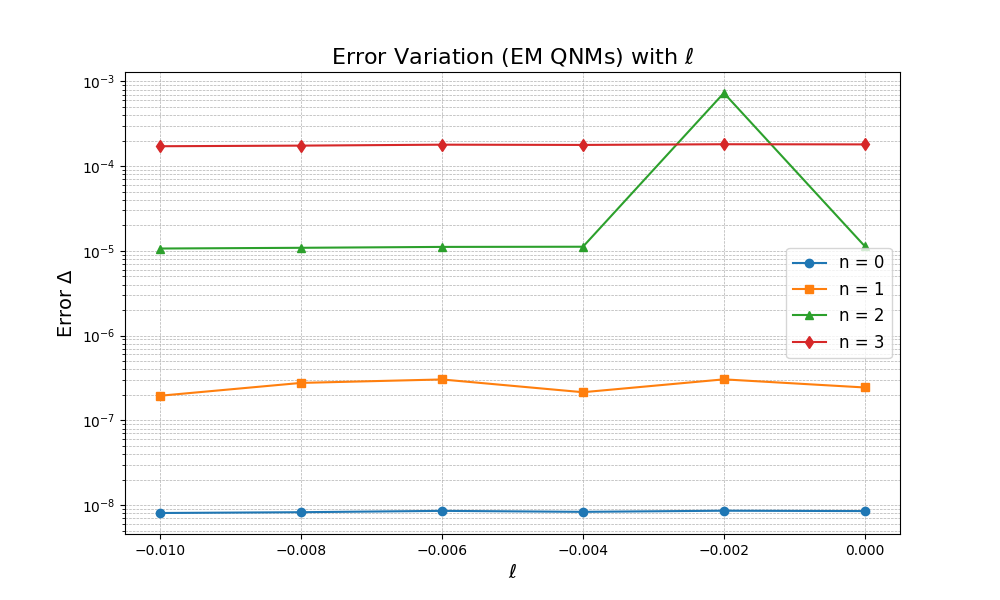}
    \caption{Errors in Table \ref{tab:emqnmvarl}}
    \label{fig:emerrorell}
\end{figure}

The variation of the EM QNMs with $\ell$ is shown in the data listed in Table \ref{tab:emqnmvarl} and visualized in Fig. \ref{fig:emqnmvarell}. As shown in the figure, as $\ell$ increases from $-0.01$ to $0$, $\omega_{\mathrm{Re}}$ exhibits increasing behavior across all overtones. For instance, the fundamental mode increases from $\omega_{\mathrm{Re}} \approx 0.840533$ at $\ell = -0.01$ to $\omega_{\mathrm{Re}} \approx 0.853095$ at $\ell = 0$. This indicates that EM perturbations of the KR BH are weakly sensitive to $\ell$. Next, $\omega_{\mathrm{Im}}$ becomes more negative as $\ell$ increases, though the changes are subtle. For the fundamental mode, $\omega_{\mathrm{Im}}$ changes from approximately $-0.0939104$ at $\ell = -0.01$ to $-0.0958599$ at $\ell = 0$. A similar trend is observed for the higher overtones. This indicates that the damping of the EM QNMs becomes slightly more efficient as $\ell$ increases. Physically, this may reflect a modification of the effective potential barrier, leading to enhanced leakage of energy from the perturbation field. As shown in Fig. \ref{fig:emerrorell}, for the lower overtones ($n=0$ and $n=1$), the errors remain significantly small (on the order of $10^{-9}$ and $10^{-7}$, respectively), suggesting robust numerical convergence of the WKB method used. However, for the higher overtones ($n=2$ and $n=3$), certain values of $\ell$ exhibit larger errors. Notably, for $n=2$ the error at $\ell = -0.002$ is an outlier, and a similar anomaly is observed for $n=3$ at $\ell = -0.004$. These deviations could indicate increased sensitivity of the numerical method in the parameter space or potentially signal more intricate physical behavior in the effective potential near these parameter values.

\begin{table}[htbp]
\centering
\begin{tabular}{|c|c|c|c|}
    \hline
    $\ell$ & $n$ & $\omega$  & $\Delta$ \\
    \hline
    -0.01 & 0 & $0.840533\, -0.0939104 i$ & $8.05845\times10^{-9}$ \\
        & 1 & $0.828964\, -0.283425 i$ & $1.95094\times10^{-7}$ \\
        & 2 & $0.806921\, -0.477888 i$ & $0.000010659$ \\
        & 3 & $0.77663\, -0.680049 i$ & $0.000171779$ \\
    \hline
    -0.008 & 0 & $0.84302\, -0.0942957 i$ & $8.23541\times10^{-9}$ \\
        & 1 & $0.831401\, -0.284589 i$ & $2.76522\times10^{-7}$ \\
        & 2 & $0.809259\, -0.479854 i$ & $0.0000108714$ \\
        & 3 & $0.778834\, -0.682856 i$ & $0.000174666$ \\
  \hline
    -0.006 & 0 & $0.84552\, -0.0946833 $i & $8.53950\times10^{-9}$ \\
        & 1 & $0.833849\, -0.28576 i$ & $3.03650\times10^{-7}$ \\
        & 2 & $0.811609\, -0.481832 i$ & $0.0000111374$ \\
        & 3 & $0.781051\, -0.68568 i$ & $0.000179202$ \\
    \hline
    -0.004 & 0 & $0.848033\, -0.0950732 i$ & $8.31609\times10^{-9}$ \\
        & 1 & $0.836309\, -0.286938 i$ & $2.14644\times10^{-7}$ \\
        & 2 & $0.813971\, -0.483824 i$ & $0.0000111912$ \\
        & 3 & $0.783283\, -0.688529 i$ & $0.000177966$ \\
    \hline
    -0.002 & 0 & $0.850558\, -0.0954654 i$ & $8.57299\times10^{-9}$ \\
        & 1 & $0.838782\, -0.288123 i$ & $3.04502\times10^{-7}$ \\
        & 2 & $0.816688\, -0.485716 i$ & $0.000729912$ \\
        & 3 & $0.785518\, -0.691381 i$ & $0.000181332$ \\
    \hline
    0 & 0 & $0.853095\, -0.0958599 i$ & $8.48987\times10^{-9}$ \\
        & 1 & $0.841267\, -0.289315 i$ & $2.44188\times10^{-7}$ \\
        & 2 & $0.818728\, -0.48784 i$ & $0.0000113506$ \\
        & 3 & $0.787769\, -0.694263 i$ & $0.000180529$ \\
    \hline
\end{tabular}
\caption{Variation of $l=4$ EM QNMs with $\ell$ and fixed $\eta =0.3$}
\label{tab:emqnmvarl}
\end{table}

\subsubsection{Variation of EM QNMs with $\eta$}
Here, we examine the variation of the EM QNM frequencies of the KR black hole as a function of the model parameter $\eta$, while keeping the multipole parameter fixed at $\ell=-0.01$. The real part of the EM QNM frequencies, $\omega_{\mathrm{Re}}$, exhibits systematic variation with $\eta$. For the fundamental mode, $\omega_{\mathrm{Re}}$ remains nearly constant, with a value of approximately $0.84053$. The higher overtones ($n=1,2,3$) also exhibit subtle variations in $\omega_{\mathrm{Re}}$ as $\eta$ increases from $0$ to $0.9$. These trends suggest that EM perturbations are weakly sensitive to the parameter $\eta$ when $\ell$ is fixed. Thus, the oscillation frequencies can be considered robust against variations in $\eta$. In contrast to the real part, $\omega_{\mathrm{Im}}$ exhibits pronounced variations with $\eta$. For $n=0$, $\omega_{\mathrm{Im}}$ changes slightly from $-0.09397$ at $\eta=0$ to approximately $-0.09340$ at $\eta=0.9$, indicating a minor reduction in the damping rate. Similar behavior is observed for the higher overtones. Notably, although the absolute variations are small, the consistent decreasing trend of $|\omega_{\mathrm{Im}}|$ with increasing $\eta$ suggests that the damping of the EM QNMs is marginally reduced at higher values of the model parameter, which may be interpreted as a weakening of the dissipative mechanism, possibly due to subtle alterations in the effective potential's barrier height or width, which governs the leakage of energy from the perturbation.
The errors in the fundamental mode ($n=0$) are significantly small, indicating the high accuracy of the WKB method used. Similarly, the error estimates for the $n=1$ mode remain on the order of $10^{-7}$, except for minor fluctuations. For the higher overtones ($n=2$ and $n=3$), the errors are slightly larger and exhibit variability. For instance, for the error corresponding to $n=2$ is on the order of $10^{-5}$ to $10^{-4}$, while for $n=3$ it ranges approximately between $10^{-4}$ and $10^{-3}$. Some outliers are also noted (for example, at $\eta=0.5$ for $n=2$ or at $\eta=0.6$ for $n=3$). These anomalies might indicate either increased numerical sensitivity in the computation of higher overtones or possibly the emergence of resonant behavior.

\begin{table}[htbp]
\centering
\begin{tabular}{|c|c|c|c|}
    \hline
    $\eta$ & $n$ & $\omega$  & $\Delta$ \\
    \hline
    0 & 0 & $0.840531\, -0.0939748 i$ & $8.42754\times10^{-8}$ \\
        & 1 & $0.828992\, -0.283609 i$ & $8.80875\times10^{-7}$ \\
        & 2 & $0.806999\, -0.47817 i$ & $0.0000135377$ \\
        & 3 & $0.776772\, -0.680406 i$ & $0.000175467$ \\
    \hline
    0.1 & 0 & $0.840531\, -0.0939675 i$ & $8.27932\times10^{-9}$ \\
        & 1 & $0.828989\, -0.283588 i$ & $2.86915\times10^{-7}$ \\
        & 2 & $0.806968\, -0.478154 i$ & $0.000047848$ \\
        & 3 & $0.776717\, -0.680383 i$ & $0.000133298$ \\
    \hline
    0.2 & 0 & $0.840531\, -0.0939461 i$ & $8.21734\times10^{-9}$ \\
        & 1 & $0.828979\, -0.283527 i$ & $2.81217\times10^{-7}$ \\
        & 2 & $0.806961\, -0.478041 i$ & $0.000010762$ \\
        & 3 & $0.776692\, -0.680243 i$ & $0.000169717$ \\
    \hline
    0.3 & 0 & $0.840533\, -0.0939104 i$ & $8.05845\times10^{-9}$ \\
        & 1 & $0.828964\, -0.283425 i$ & $1.95094\times10^{-7}$ \\
        & 2 & $0.806921\, -0.477888 i$ & $0.000010659$ \\
        & 3 & $0.77663\, -0.680049 i$ & $0.000171779$ \\
    \hline
    0.4 & 0 & $0.840535\, -0.0938604 i$ & $7.81494\times10^{-9}$ \\
        & 1 & $0.828944\, -0.283283 i$ & $1.53101\times10^{-7}$ \\
        & 2 & $0.806864\, -0.477673 i$ & $0.0000102767$ \\
        & 3 & $0.776538\, -0.67979 i$ & $0.000160844$ \\
    \hline
    0.5 & 0 & $0.840539\, -0.0937962 i$ & $6.91838\times10^{-9}$ \\
        & 1 & $0.828918\, -0.283101 i$ & $4.79139\times10^{-7}$ \\
        & 2 & $0.806761\, -0.477425 i$ & $0.0000876684$ \\
        & 3 & $0.776197\, -0.679699 i$ & $0.000228446$ \\
    \hline
    0.6 & 0 & $0.840545\, -0.0937176 i$ & $8.08020\times10^{-9}$ \\
        & 1 & $0.828888\, -0.282877 i$ & $2.54925\times10^{-7}$ \\
        & 2 & $0.806704\, -0.477058 i$ & $0.0000105373$ \\
        & 3 & $0.78497\, -0.680625 i$ & $0.0177751$ \\
    \hline
    0.7 & 0 & $0.840555\, -0.0936248 $i & $8.29196\times10^{-9}$ \\
        & 1 & $0.828854\, -0.282613 i$ & $2.94306\times10^{-7}$ \\
        & 2 & $0.806841\, -0.476598 i$ & $0.000498975$ \\
        & 3 & $0.776106\, -0.678519 i$ & $0.000171347$ \\
    \hline
    0.8 & 0 & $0.840569\, -0.0935177 i$ & $8.11456\times10^{-9}$ \\
        & 1 & $0.828817\, -0.282309 i$ & $2.65578\times10^{-7}$ \\
        & 2 & $0.806686\, -0.476132 i$ & $0.000427405$ \\
        & 3 & $0.775912\, -0.677949 i$ & $0.000176698$ \\
    \hline
    0.9 & 0 & $0.840589\, -0.0933962 i$ & $1.45159\times10^{-7}$ \\
        & 1 & $0.828778\, -0.281963 i$ & $2.54807\times10^{-6}$ \\
        & 2 & $0.806333\, -0.475672 i$ & $0.0000144084$ \\
        & 3 & $0.77553\, -0.677351 i$ & $0.000102768$ \\
    \hline
\end{tabular}
\caption{Variation of $l=4$ EM QNMs with $\eta$ and fixed $\ell =-0.01$}
\label{tab:emqnmvareta}
\end{table}

\begin{figure}[htbp]
     \centering
     \begin{subfigure}[b]{\textwidth}
         \centering
         \includegraphics[width=\textwidth]{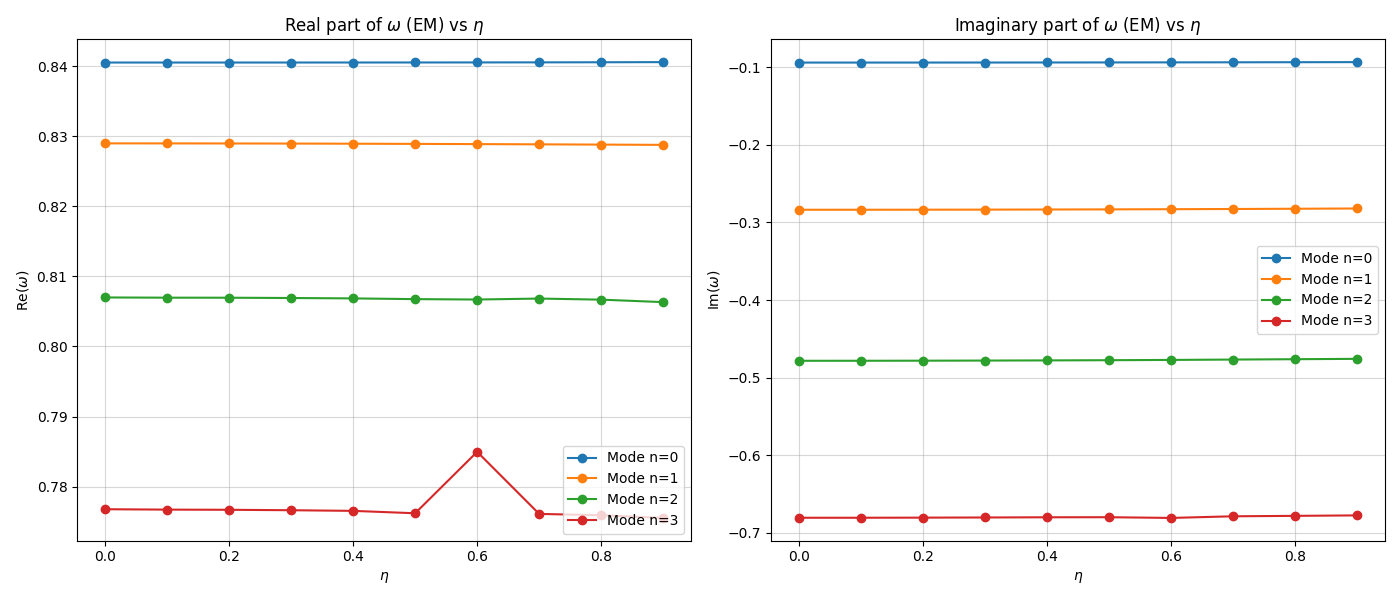}
         %\caption{Variation with $\ell$}
         %\label{fig:potvarl}
     \end{subfigure}
     %\hfill
     \begin{subfigure}[b]{\textwidth}
         \centering
         \includegraphics[width=\textwidth]{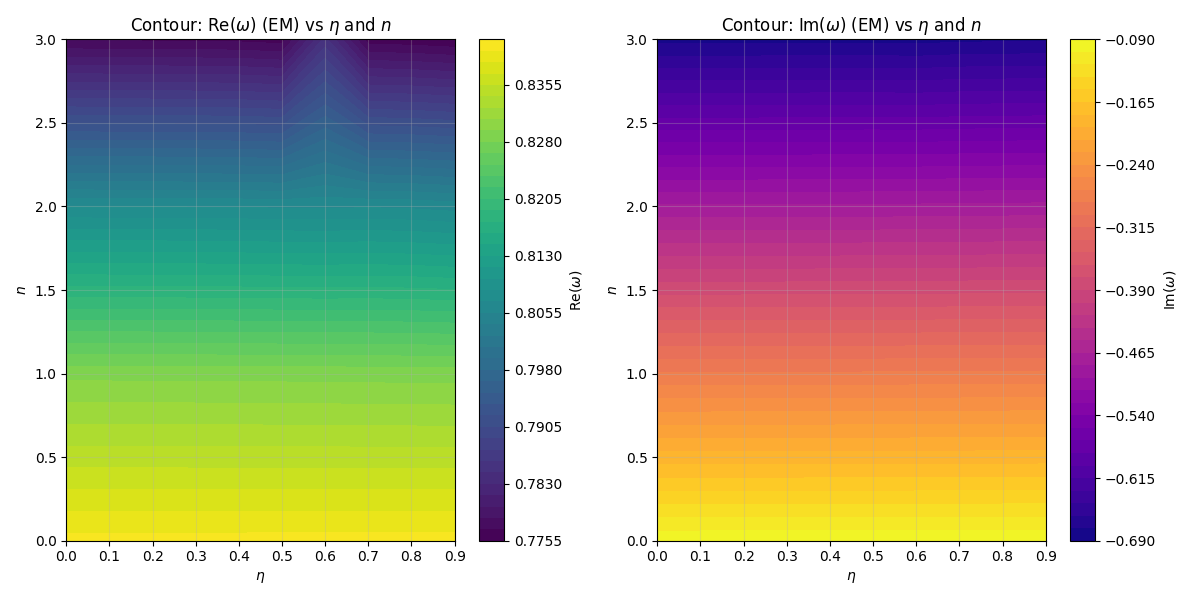}
         %\caption{Variation with $\eta$}
         %\label{fig:potvareta}
     \end{subfigure}
    \caption{Variation of the $l=4$ EM QNMs with $\eta$ and fixed $\ell = -0.01$.}
    \label{fig:emqnmvareta}
\end{figure}

\begin{figure}
    \centering
    \includegraphics[width=0.7\textwidth]{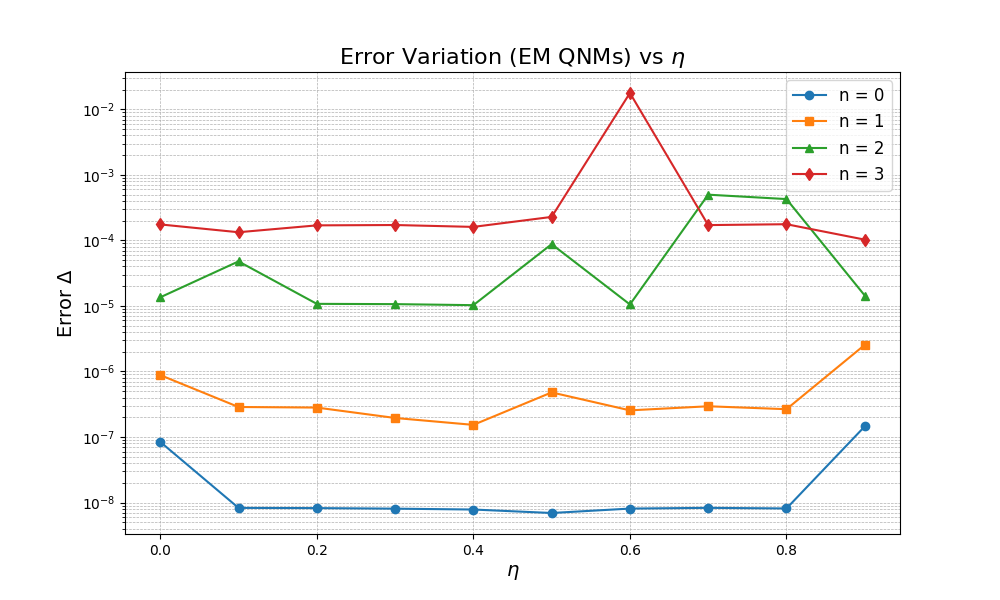}
    \caption{Errors in Table \ref{tab:emqnmvareta}}
    \label{fig:emerroreta}
\end{figure}
%%%%%%%%%%%%%%%%%%%GRAVITATIONAL%%%%%%%%%%%%%%%%%%%%%%%
\subsection{Gravitational Perturbations}
To evaluate axial gravitational perturbations in effective theories such as the one in our framework, it can be considered that the black hole is described by Einstein gravity minimally coupled to an anisotropic source, and perturbations can be considered as encoded in the perturbations of the gravitational field equation and the corresponding anisotropic energy--momentum tensor. In the tetrad formalism, the axial components of perturbed energy--momentum tensor are zero \cite{Chen:2019iuo}, and thus, the master equation can be derived from $R_{(a)(b)} = 0$; the $\theta \, \phi$ and $r\, \phi$ components yield \cite{bouhmadi2020consistent}
\begin{align}
\left[ r^2 \sqrt{|g_{tt}| g_{rr}^{-1}}\, (b_{,\theta} - c_{,r}) \right]_{,r} = r^2 \sqrt{|g_{tt}|^{-1} g_{rr}}\, (a_{,\theta} - c_{,t})_{,t}, \label{g1}\\
\left[ r^2 \sqrt{|g_{tt}| g_{rr}^{-1}}\, (c_{,r} - b_{,\theta}) \sin^3\theta  \right]_{,\theta} = \dfrac{r^4 \sin^3\theta}{\sqrt{|g_{tt}| g_{rr}}}\, (a_{,r} - b_{,t})_{,t}. \label{g2}
\end{align}

\begin{figure}[htbp]
     \centering
     \begin{subfigure}[t]{0.49\textwidth}
         \centering
         \includegraphics[width=\textwidth]{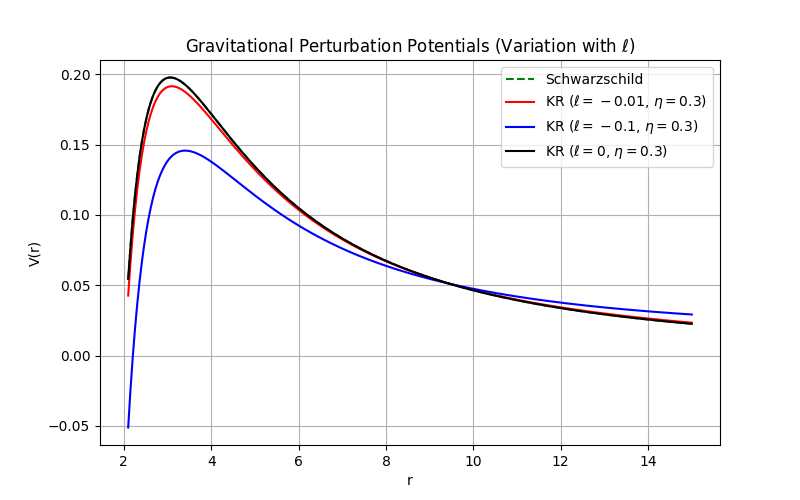}
         \caption{Variation with $\ell$}
         \label{fig:gravpotvarl}
     \end{subfigure}
     %\hfill
     \begin{subfigure}[t]{0.49\textwidth}
         \centering
         \includegraphics[width=\textwidth]{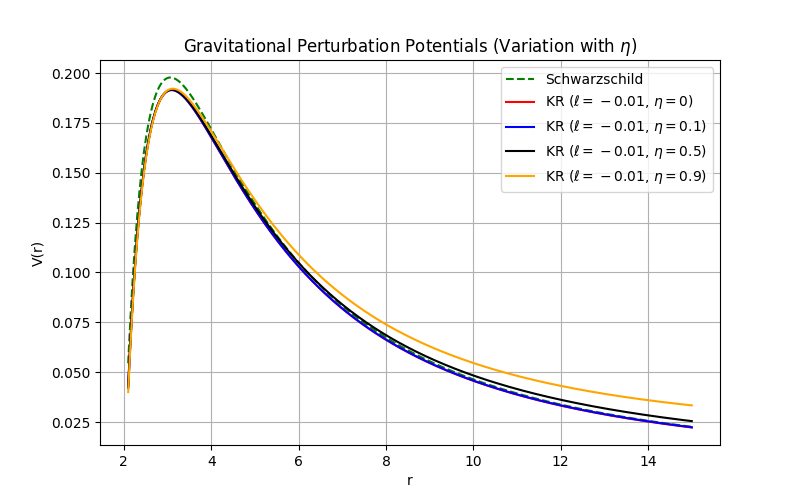}
         \caption{Variation with $\eta$}
         \label{fig:gravpotvareta}
     \end{subfigure}
    \caption{Variation of the gravitational perturbation potential with $\ell$ and $\eta$.}
    \label{fig:gravpotvar}
\end{figure}

Now, considering the ansatz $\mathcal{F}_g (r, \theta) = \mathcal{F}_g (r) Y(\theta)$, using the redefinition $\psi_g r = \mathcal{F}_g$, and introducing the tortise coordinate defined earlier, the master perturbation equation can be derived from Eqs. \eqref{g1} and \eqref{g2} as
\begin{equation}
\partial^2_{r_*} \psi_g + \omega^2 \psi_g = V_g(r) \psi_g,
\end{equation}

where the effective potential is given by
\begin{equation}\label{Vg}
V_g(r) = |g_{tt}| \left[ \dfrac{2}{r^2} \left( \dfrac{1}{g_{rr}} - 1 \right) + \dfrac{l(l+1)}{r^2} - \dfrac{1}{r \sqrt{|g_{tt}| g_{rr}}} \left( \dfrac{d}{dr} \sqrt{|g_{tt}| g_{rr}^{-1}} \right) \right].
\end{equation}
%%% PROF. DESHAMUKHYA TO DETAIL THIS PART'S RESULTS%%%
The dependence of the gravitational perturbation potential on the model parameters $\ell$ and $\eta$ is shown in Fig. \ref{fig:gravpotvar}

\subsubsection{Variation of gravitational QNMs with $\ell$}
\begin{table}[htbp]
\centering
\begin{tabular}{|c|c|c|c|}
    \hline
    $\ell$ & $n$ & $\omega$  & $\Delta$ \\
    \hline
    -0.01 & 0 & $0.826017\, -0.0933595 i$ & $1.60512\times10^{-8}$ \\
        & 1 & $0.814202\, -0.281814 i$ & $3.56186\times10^{-7}$ \\
        & 2 & $0.792106\, -0.475256 i$ & $0.000852838$ \\
        & 3 & $0.760776\, -0.676758 i$ & $0.000192527$ \\
    \hline
    -0.008 & 0 & $0.828511\, -0.0937451 i$ & $9.86473\times10^{-9}$ \\
        & 1 & $0.816648\, -0.282978 i$ & $1.850345\times10^{-7}$ \\
        & 2 & $0.794047\, -0.477304 i$ & $0.0000119645$ \\
        & 3 & $0.763007\, -0.679562 i$ & $0.000190059$ \\
    \hline
    -0.006 & 0 & $0.831017\, -0.094133 i$ & $1.08987\times10^{-8}$ \\
        & 1 & $0.819107\, -0.284148 i$ & $3.16565\times10^{-7}$ \\
        & 2 & $0.796411\, -0.479276 i$ & $0.0000118237$ \\
        & 3 & $0.765234\, -0.68236 i$ & $0.00018886$ \\
    \hline
    -0.004 & 0 & $0.833537\, -0.0945231 i$ & $9.92957\times10^{-9}$ \\
        & 1 & $0.821578\, -0.285326 i$ & $1.64192\times10^{-7}$ \\
        & 2 & $0.79879\, -0.481264 i$ & $0.0000121857$ \\
        & 3 & $0.767494\, -0.685205 i$ & $0.000192962$ \\
    \hline
    -0.002 & 0 & $0.836069\, -0.0949156 i$ & $2.03060\times10^{-8}$ \\
        & 1 & $0.824061\, -0.286509 i$ & $5.48275\times10^{-7}$ \\
        & 2 & $0.801176\, -0.483257 i$ & $0.0000132747$ \\
        & 3 & $0.769734\, -0.688039 i$ & $0.000205674$ \\
    \hline
    0 & 0 & $0.838615\, -0.0953104 i$ & $1.10228\times10^{-8}$ \\
        & 1 & $0.826557\, -0.287701 i$ & $2.51917\times10^{-7}$ \\
        & 2 & $0.80358\, -0.485269 i$ & $0.0000123588$ \\
        & 3 & $0.772017\, -0.690916 i$ & $0.000192333$ \\
    \hline
\end{tabular}
\caption{Variation of $l=4$ gravitational QNMs with $\ell$ and fixed $\eta =0.3$}
\label{tab:gravqnmvarl}
\end{table}

\begin{figure}[htbp]
     \centering
     \begin{subfigure}[b]{\textwidth}
         \centering
         \includegraphics[width=\textwidth]{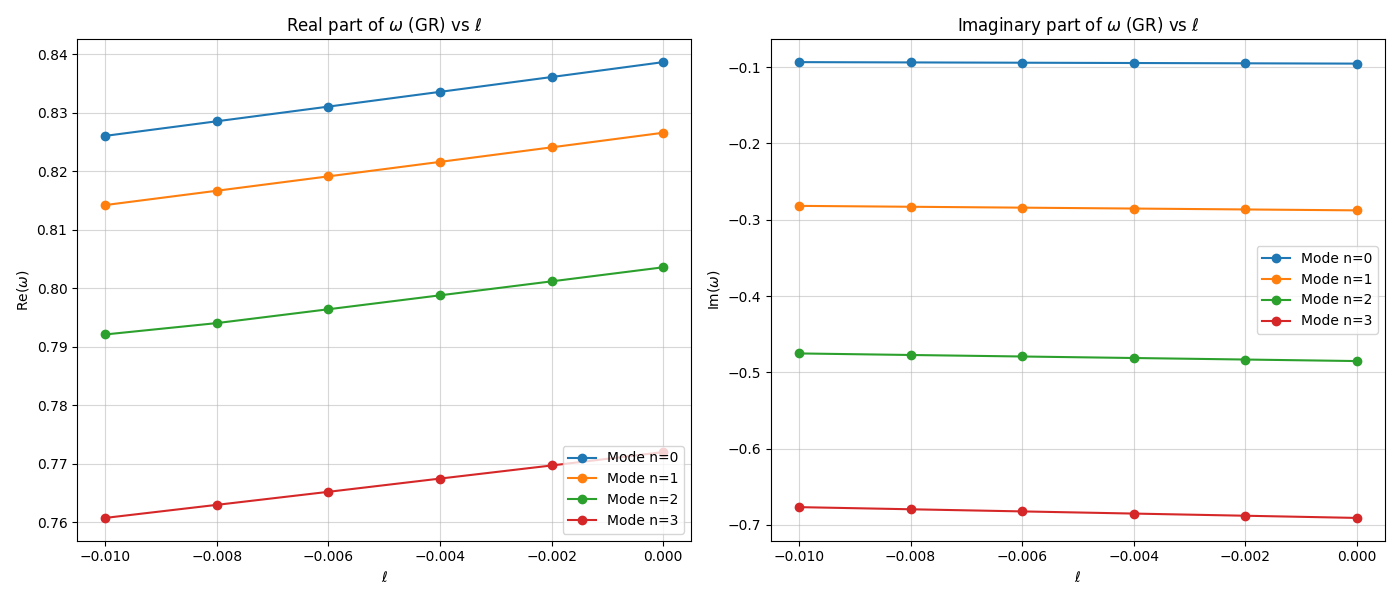}
     \end{subfigure}
     %\hfill
     \begin{subfigure}[b]{\textwidth}
         \centering
         \includegraphics[width=\textwidth]{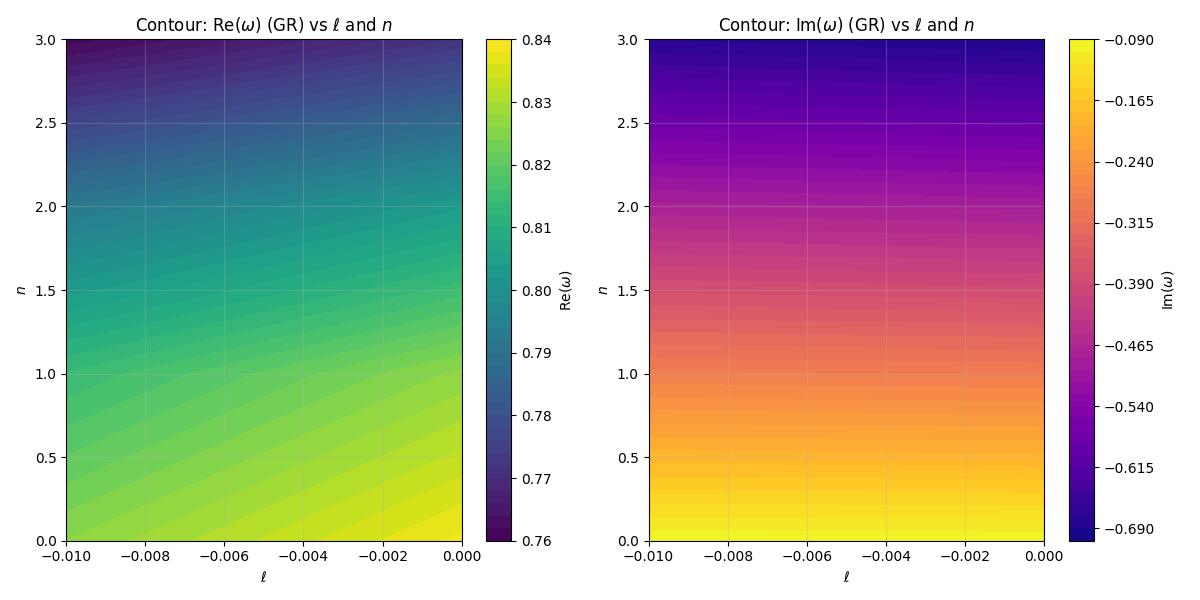}
     \end{subfigure}
    \caption{Variation of the $l=4$ gravitational QNMs with $\ell$ and fixed $\eta = 0.3$.}
    \label{fig:qnmvarellgrav}
\end{figure}

This section investigates the variation of the gravitational QNMs of the KR black hole with respect to the parameter $\ell$, while keeping the model parameter fixed at $\eta=0.3$. The estimated frequency-domain data are presented in Table \ref{tab:gravqnmvarl} and visualized in Fig. \ref{fig:qnmvarellgrav}. It can be seen that $\omega_{\mathrm{Re}}$ exhibits an increasing trend with $\ell$ for all overtones. For instance, in the fundamental mode ($n=0$), $\omega_{\mathrm{Re}}$ increases from approximately $0.8260$ at $\ell=-0.01$ to $0.8386$ at $\ell=0$. This trend is consistent across higher overtones. The physical interpretation of this trend lies in the nature of the effective potential governing gravitational perturbations. Moreover, $\omega_{\mathrm{Im}}$, which determines the damping rate, also exhibits a systematic variation with $\ell$. Specifically, a steady increase in the magnitude of $\omega_{\mathrm{Im}}$ (that is, it becomes more negative) is observed with increasing $\ell$. For the fundamental mode, $\omega_{\mathrm{Im}}$ changes from approximately $-0.09336$ at $\ell=-0.01$ to $-0.09531$ at $\ell=0$. The trend is more pronounced for higher overtones. This behavior suggests that perturbations decay more rapidly for larger values of $\ell$, implying enhanced damping. The relative errors $\Delta$ corresponding to the estimated QNMs are generally small, indicating good numerical convergence. For the fundamental mode, the errors remain on the order of $10^{-8}$ to $10^{-9}$. For $n=1$, the errors are slightly larger (on the order of $10^{-7}$), but within acceptable limits. For the higher overtones ($n=2,3$), the errors show higher variability. Notably, for $n=2$, the error at $\ell=-0.01$ is significantly larger ($\sim 0.00085$), while for other $\ell$ values, it remains on the order of $10^{-5}$ to $10^{-6}$. This suggests that numerical accuracy is sensitive for higher overtones.

\begin{figure}
    \centering
    \includegraphics[width=0.7\textwidth]{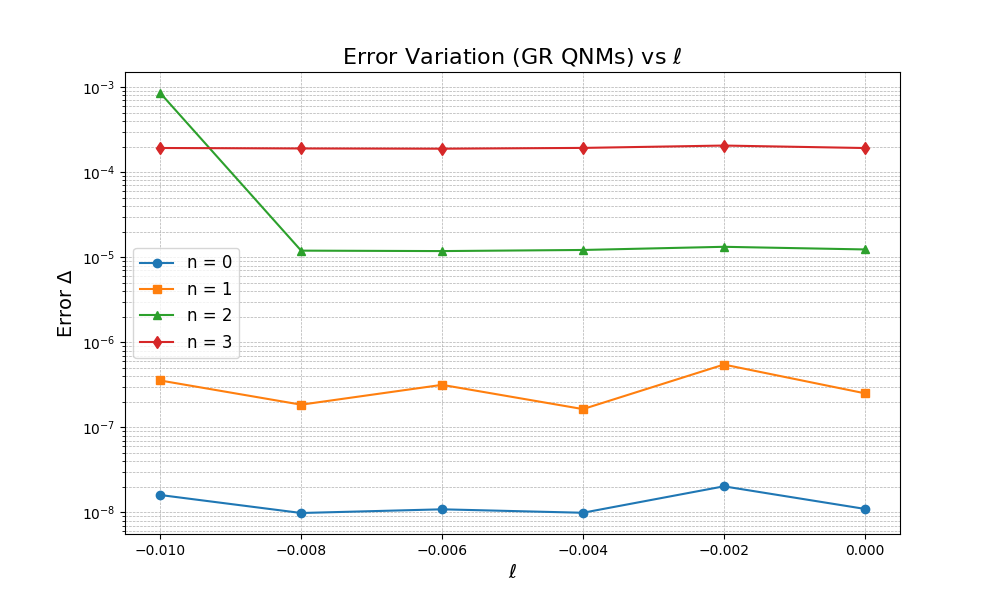}
    \caption{Errors in Table \ref{tab:gravqnmvarl}}
    \label{fig:errorellgrav}
\end{figure}

\subsubsection{Variation of gravitational QNMs with $\eta$}
\begin{table}[htbp]
\centering
\begin{tabular}{|c|c|c|c|}
    \hline
    $\eta$ & $n$ & $\omega$  & $\Delta$ \\
    \hline
    0 & 0 & $0.825977\, -0.0934407 i$ & $7.98485\times10^{-8}$ \\
        & 1 & $0.81421\, -0.282043 i$ & $1.77510\times10^{-6}$ \\
        & 2 & $0.79179\, -0.475685 i$ & $0.0000273425$ \\
        & 3 & $0.761013\, -0.677201 i$ & $0.000235264$ \\
    \hline
    0.1 & 0 & $0.825981\, -0.0934315 i$ & $9.83194\times10^{-9}$ \\
        & 1 & $0.81421\, -0.282016 i$ & $1.97713\times10^{-7}$ \\
        & 2 & $0.791769\, -0.475636 i$ & $0.0000118918$ \\
        & 3 & $0.760921\, -0.677105 i$ & $0.000194114$ \\
    \hline
    0.2 & 0 & $0.825994\, -0.0934046 i$ & $7.02244\times10^{-9}$ \\
        & 1 & $0.814207\, -0.281941 i$ & $1.28577\times10^{-7}$ \\
        & 2 & $0.791742\, -0.475527 i$ & $0.0000121541$ \\
        & 3 & $0.760877\, -0.676987 i$ & $0.00018912$ \\
    \hline
    0.3 & 0 & $0.826017\, -0.0933595 i$ & $1.60512\times10^{-8}$ \\
        & 1 & $0.814202\, -0.281814 i$ & $3.56186\times10^{-7}$ \\
        & 2 & $0.792106\, -0.475256 i$ & $0.000852838$ \\
        & 3 & $0.760776\, -0.676758 i$ & $0.000192527$ \\
    \hline
    0.4 & 0 & $0.826049\, -0.0932963 i$ & $9.70264\times10^{-9}$ \\
        & 1 & $0.814196\, -0.281638 i$ & $1.77838\times10^{-7}$ \\
        & 2 & $0.791628\, -0.475085 i$ & $0.0000117009$ \\
        & 3 & $0.760658\, -0.67647 i$ & $0.000180624$ \\
    \hline
    0.5 & 0 & $0.826091\, -0.0932151 i$ & $2.47987\times10^{-8}$ \\
        & 1 & $0.814188\, -0.281413 i$ & $7.26756\times10^{-7}$ \\
        & 2 & $0.791552\, -0.474761 i$ & $0.0000172737$ \\
        & 3 & $0.760553\, -0.67609 i$ & $0.000210213$ \\
    \hline
    0.6 & 0 & $0.826144\, -0.0931154 i$ & $1.10924\times10^{-8}$ \\
        & 1 & $0.814181\, -0.281134 i$ & $2.68198\times10^{-7}$ \\
        & 2 & $0.791441\, -0.474351 i$ & $0.0000113473$ \\
        & 3 & $0.760305\, -0.675597 i$ & $0.000181959$ \\
    \hline
    0.7 & 0 & $0.82621\, -0.0929975 i$ & $1.02441\times10^{-8}$ \\
        & 1 & $0.814173\, -0.280806 i$ & $2.01366\times10^{-7}$ \\
        & 2 & $0.791322\, -0.473876 i$ & $0.0000107261$ \\
        & 3 & $0.760081\, -0.675046 i$ & $0.000169263$ \\
    \hline
    0.8 & 0 & $0.826289\, -0.0928612 i$ & $9.58757\times10^{-9}$ \\
        & 1 & $0.814167\, -0.280428 i$ & $1.09999\times10^{-7}$ \\
        & 2 & $0.791187\, -0.473329 i$ & $0.0000109236$ \\
        & 3 & $0.759823\, -0.674407 i$ & $0.000167509$ \\
    \hline
    0.9 & 0 & $0.826383\, -0.0927061 i$ & $1.03704\times10^{-8}$ \\
        & 1 & $0.814163\, -0.279998 i$ & $1.40012\times10^{-7}$ \\
        & 2 & $0.791032\, -0.472708 i$ & $9.98827\times10^{-6}$ \\
        & 3 & $0.759521\, -0.673683 i$ & $0.000152942$ \\
    \hline
\end{tabular}
\caption{Variation of $l=4$ gravitational QNMs with $\eta$ and fixed $\ell =-0.01$}
\label{tab:gravqnmvareta}
\end{table}

\begin{figure}[htbp]
     \centering
     \begin{subfigure}[b]{\textwidth}
         \centering
         \includegraphics[width=\textwidth]{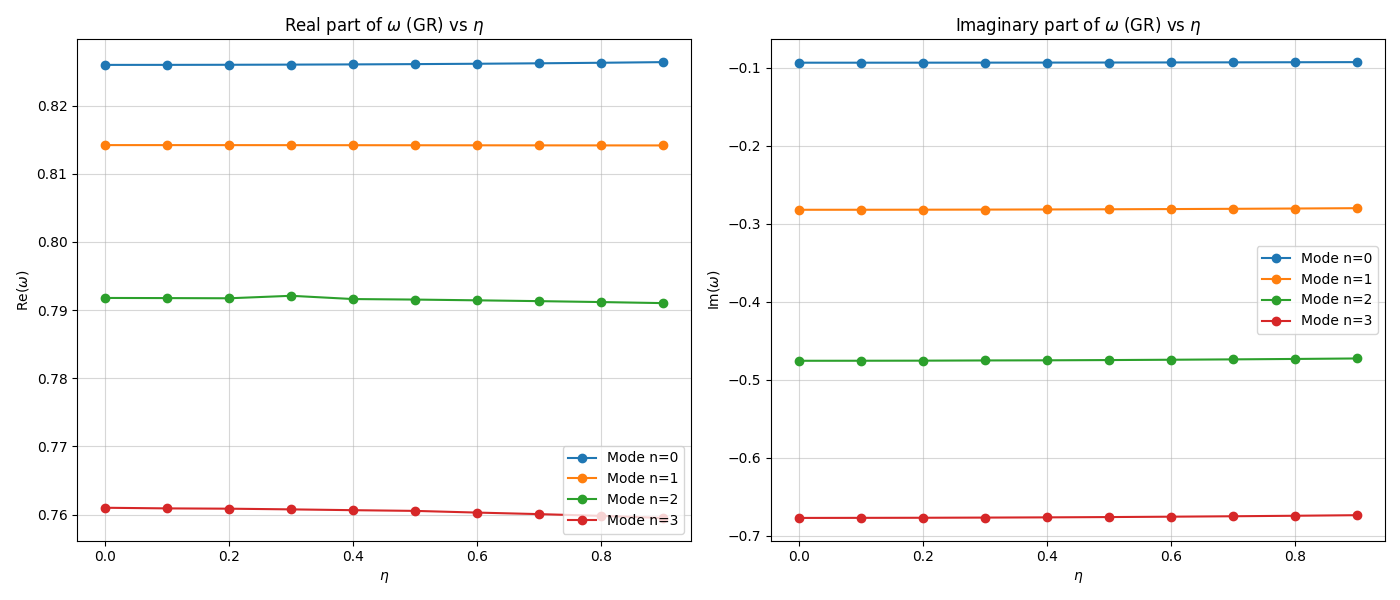}
         %\caption{Variation with $\ell$}
         %\label{fig:potvarl}
     \end{subfigure}
     %\hfill
     \begin{subfigure}[b]{\textwidth}
         \centering
         \includegraphics[width=\textwidth]{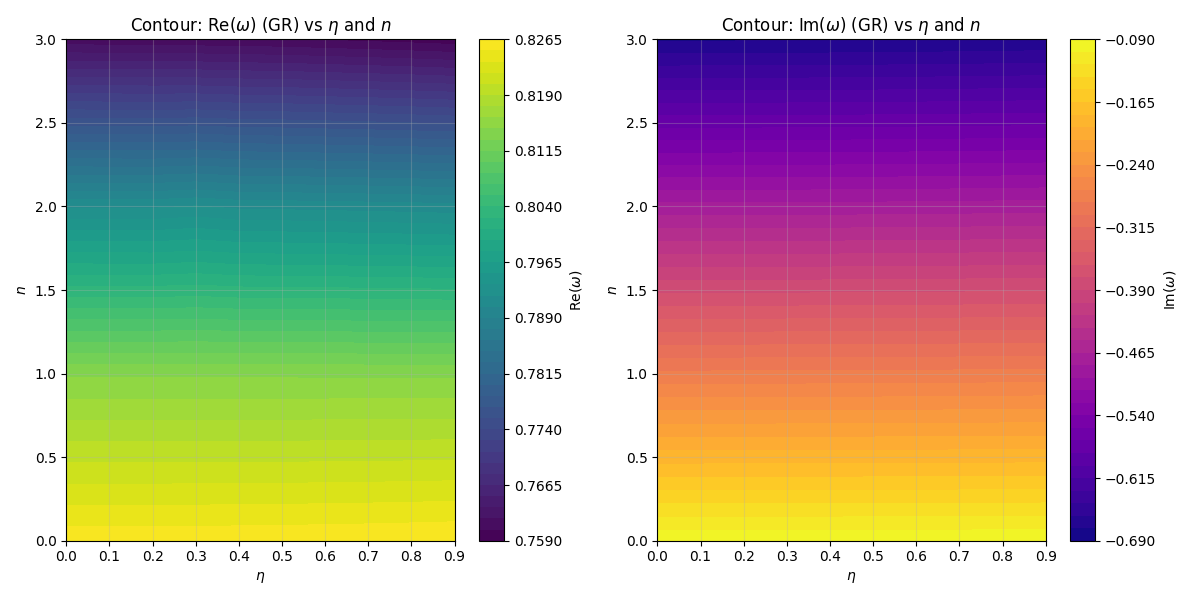}
         %\caption{Variation with $\eta$}
         %\label{fig:potvareta}
     \end{subfigure}
    \caption{Variation of the $l=4$ gravitational QNMs with $\eta$ and fixed $\ell = -0.01$.}
    \label{fig:gravqnmvareta}
\end{figure}

The estimated QNM frequencies for gravitational perturbations of the KR black hole are presented in Table \ref{tab:gravqnmvareta} and visualized in Fig. \ref{fig:gravqnmvareta}. It can be seen that $\omega_{\mathrm{Re}}$ exhibits subtle increases with $\eta$ for all overtones. For the fundamental mode, $\omega_{\mathrm{Re}}$ increases slightly from $0.82598$ at $\eta=0$ to $0.82638$ at $\eta=0.9$. This increasing trend is consistently observed for higher overtones. The weak dependence of $\omega_{\mathrm{Re}}$ on $\eta$ suggests that the oscillatory behavior of the gravitational perturbations is largely insensitive to variations in this model parameter. In contrast, $\omega_{\mathrm{Im}}$ exhibits a stronger dependence on $\eta$. The magnitude of $\omega_{\mathrm{Im}}$ (damping rate) decreases as $\eta$ increases. For $n=0$, $\omega_{\mathrm{Im}}$ increases from $-0.09344$ at $\eta=0$ to $-0.09270$ at $\eta=0.9$. This trend is consistent for the higher overtones, suggesting that perturbations decay more slowly (are long lived) as $\eta$ increases. A critical aspect of our analysis is the evaluation of the numerical errors associated with the estimated QNMs. For the fundamental mode, the errors remain significantly small ($\sim 10^{-8}$), indicating high precision. For $n=1$, the errors are slightly larger ($\sim 10^{-7}$), but remain well-controlled. The higher overtones ($n=2,3$) exhibit larger errors, especially at certain values of $\eta$. Notably, for $n=2$, the error peaks at $\eta=0.3$ with $\Delta \sim 0.00085$, while for $n=3$, fluctuations occur at multiple points, particularly near $\eta=0.5$. These anomalies may be attributed to numerical sensitivity of higher overtones. However, the overall error estimates remain within reasonable limits, supporting the reliability of the Pade-averaged WKB method.

\begin{figure}
    \centering
    \includegraphics[width=0.7\textwidth]{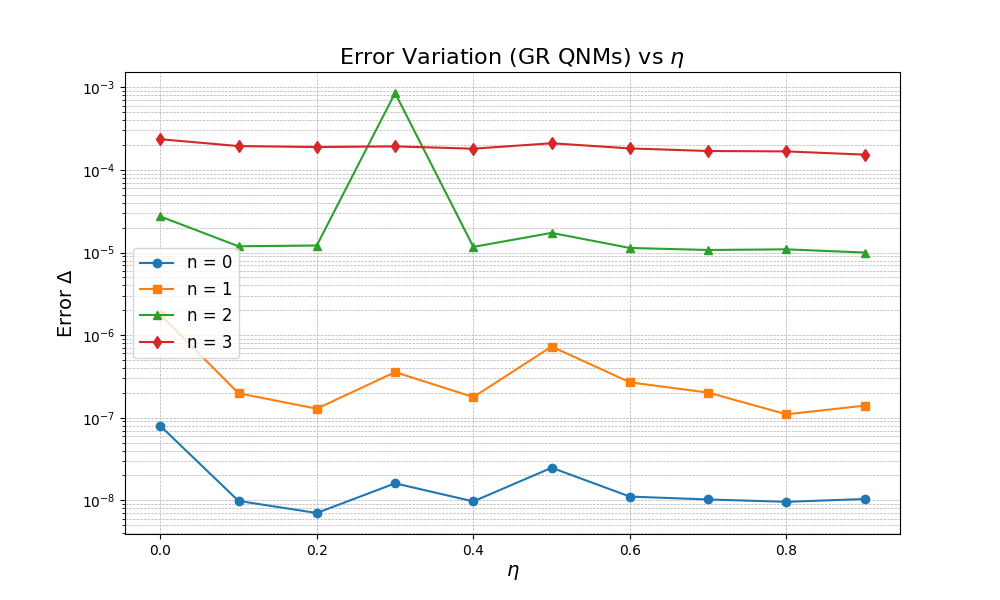}
    \caption{Errors in Table \ref{tab:gravqnmvareta}}
    \label{fig:graverroreta}
\end{figure}
\subsection{Time-domain QNMs}
The wave-like equation for perturbations can be written \textit{without} implying a stationary ansatz as \cite{konoplya2011quasinormal}:

\begin{equation}\label{eq:wavelike}
\frac{\partial^2\Phi}{\partial t^2}-\frac{\partial^2\Phi}{\partial x^2}+V(t,x)\Phi=0,
\end{equation}
Here, we denote the tortoise coordinate by $x$. The most widely discussed method to integrate the above wave equation in the time domain was developed by Gundlach, Price and Pullin \cite{Gundlach:1994, Gundlach_1994b}; we call this the GPP method. In terms of \textit{light-cone coordinates} $du = dt - dx$ and $dv = dt + dx$, the above equation can be rewritten as:

\begin{equation}\label{eq:light-cone}
\left(4\frac{\partial^2}{\partial u\partial v}+V(u,v)\right)\Phi(u,v)=0.
\end{equation}

The following discretization scheme is as per the GPP method:
\begin{eqnarray}
\Phi(N)= \Phi(W)+\Phi(E)-\Phi(S) -
\frac{h^2}{8}V(S)\left(\Phi(W)+\Phi(E)\right) + \mathcal{O}(h^4),\label{integration-scheme}
\end{eqnarray}
where we introduced letters to mark the points as follows: $S=(u,v)$, $W=(u+h,v)$, $E=(u,v+h)$, and $N=(u+h,v+h)$.

\begin{figure}[htbp]
     \centering
     \begin{subfigure}[t]{0.49\textwidth}
         \centering
         \includegraphics[width=\textwidth]{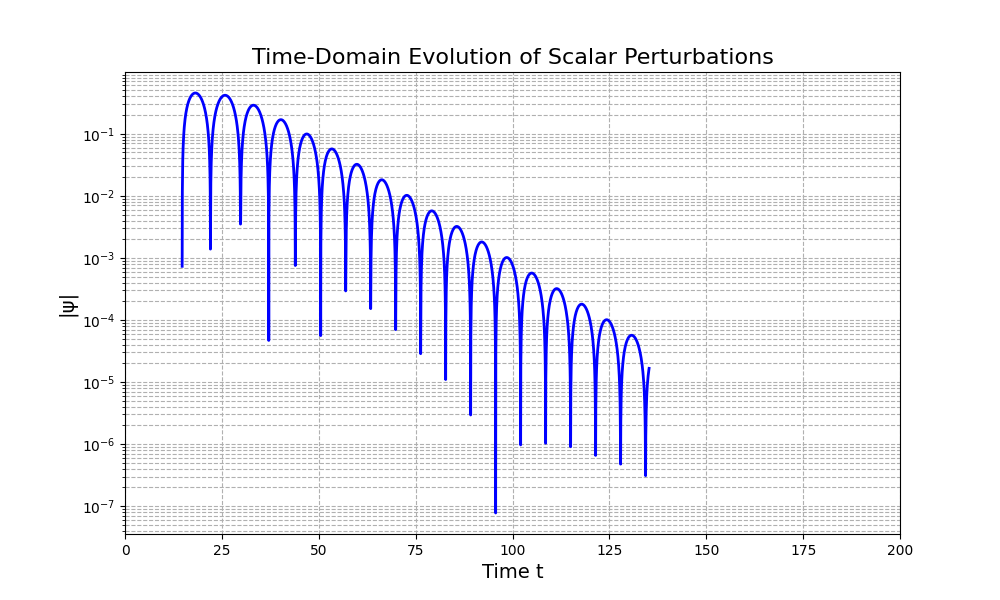}
         \caption{Time domain profile for scalar perturbations}
         \label{fig:td_scalar}
     \end{subfigure}
     \hfill
     \begin{subfigure}[t]{0.49\textwidth}
         \centering
         \includegraphics[width=\textwidth]{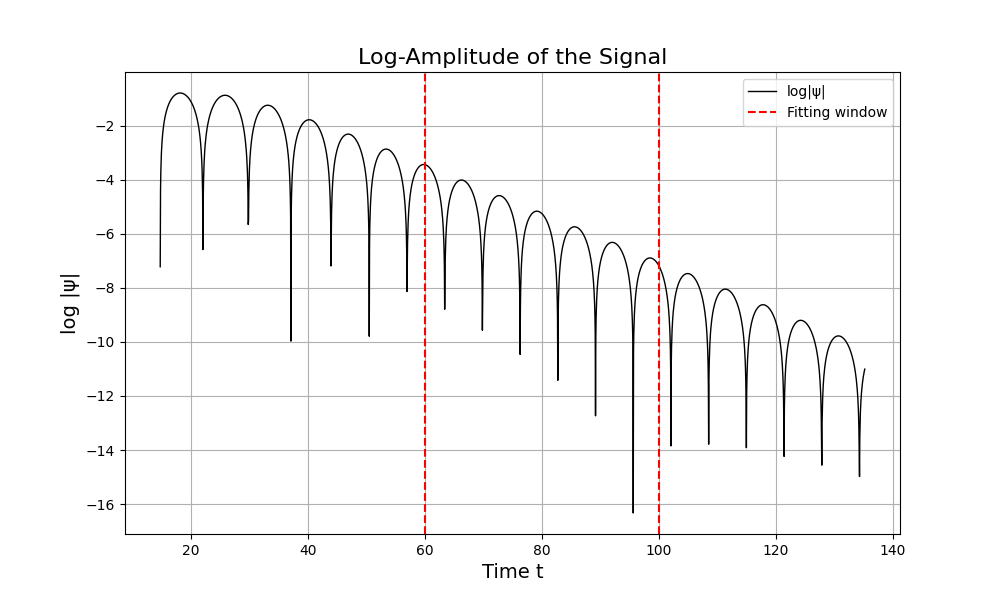}
         \caption{Log-amplitude and fitting window}
         \label{fig:td_logamp}
     \end{subfigure}
     \begin{subfigure}[t]{0.49\textwidth}
         \centering
         \includegraphics[width=\textwidth]{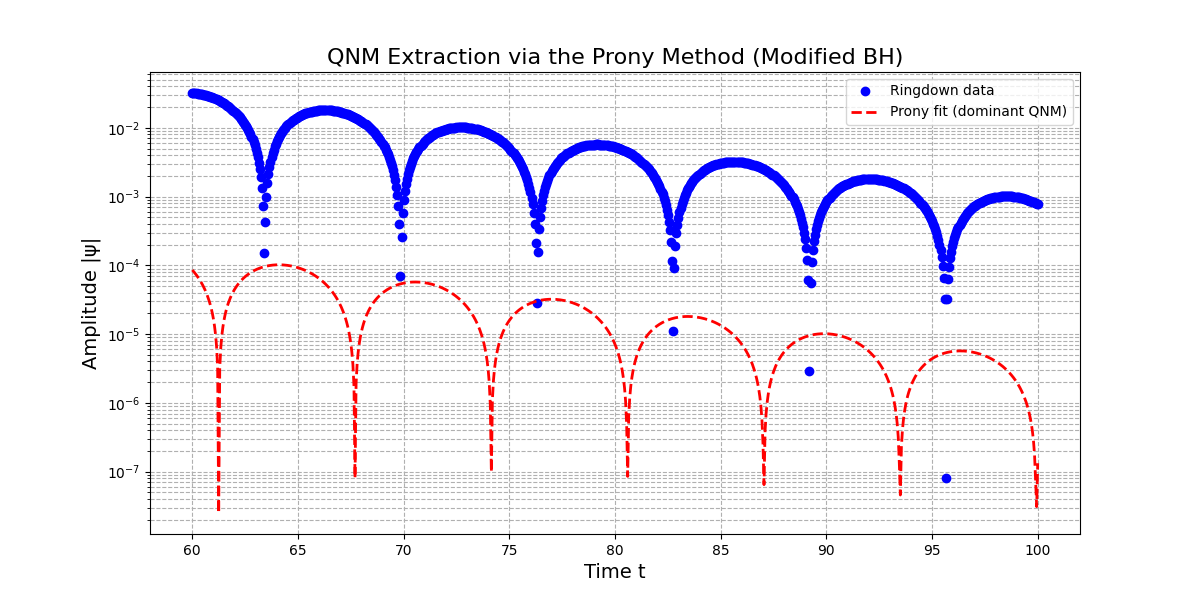}
         \caption{Prony fit of the time domain profile}
         \label{fig:prony_scalar}
     \end{subfigure}
    \caption{Time domain profiles and Prony extraction for scalar perturbations}
    \label{fig:td_scalar_full}
\end{figure}

\begin{figure}[htbp]
     \centering
     \begin{subfigure}[t]{0.49\textwidth}
         \centering
         \includegraphics[width=\textwidth]{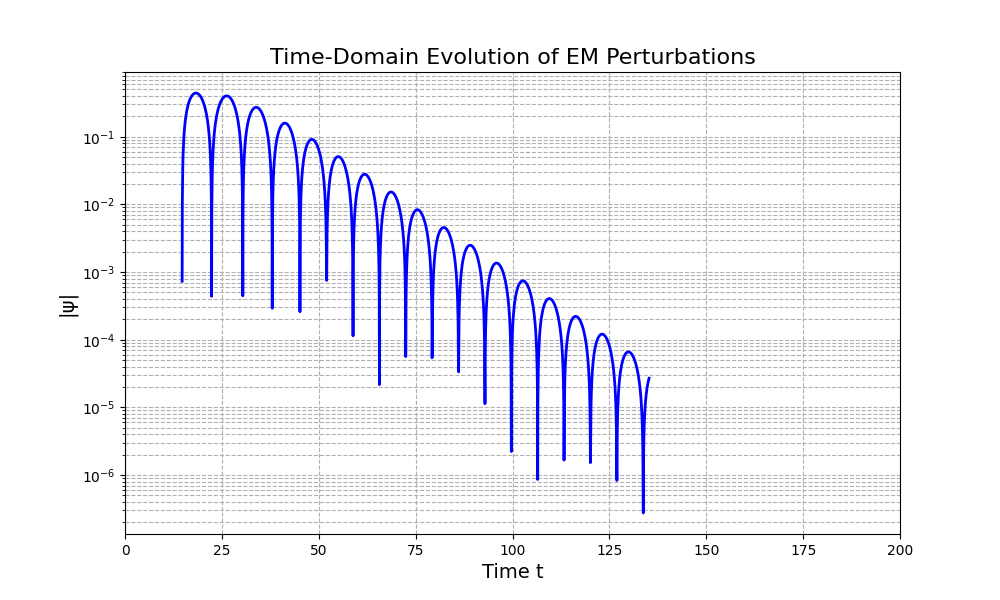}
         \caption{Time domain profile for EM perturbations}
         \label{fig:td_em}
     \end{subfigure}
     \hfill
     \begin{subfigure}[t]{0.49\textwidth}
         \centering
         \includegraphics[width=\textwidth]{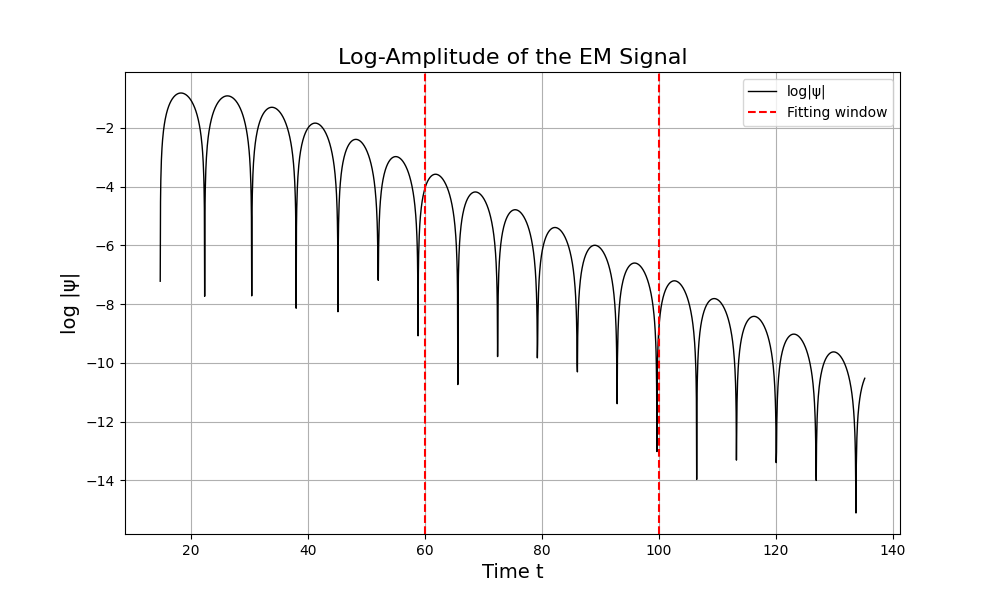}
         \caption{Log-amplitude and fitting window}
         \label{fig:td_em_logamp}
     \end{subfigure}
     \begin{subfigure}[t]{0.49\textwidth}
         \centering
         \includegraphics[width=\textwidth]{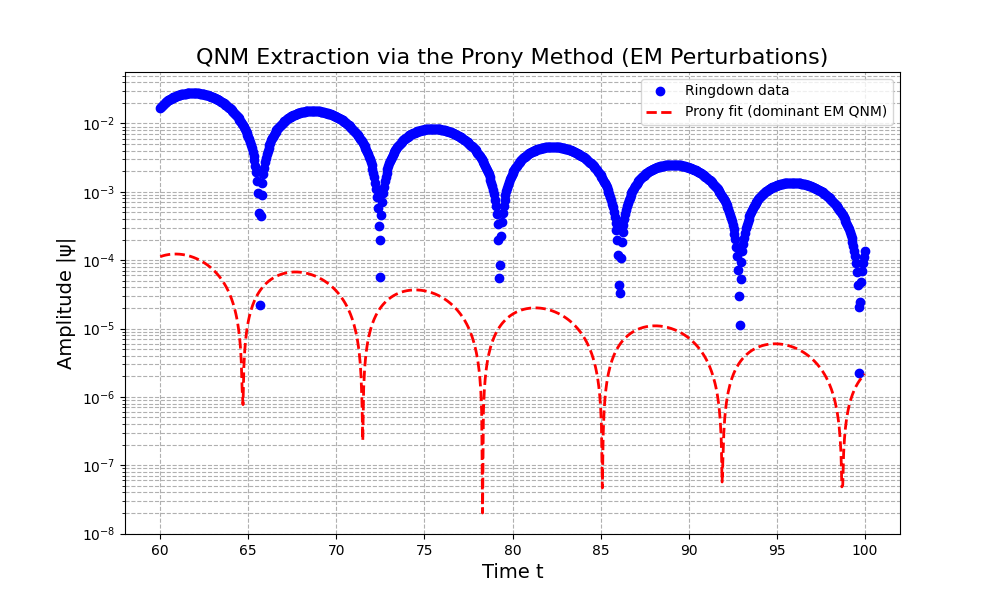}
         \caption{Prony fit of the time domain profile}
         \label{fig:prony_em}
     \end{subfigure}
    \caption{Time domain profiles and Prony extraction for EM perturbations}
    \label{fig:td_em_full}
\end{figure}

\begin{figure}[htbp]
     \centering
     \begin{subfigure}[t]{0.49\textwidth}
         \centering
         \includegraphics[width=\textwidth]{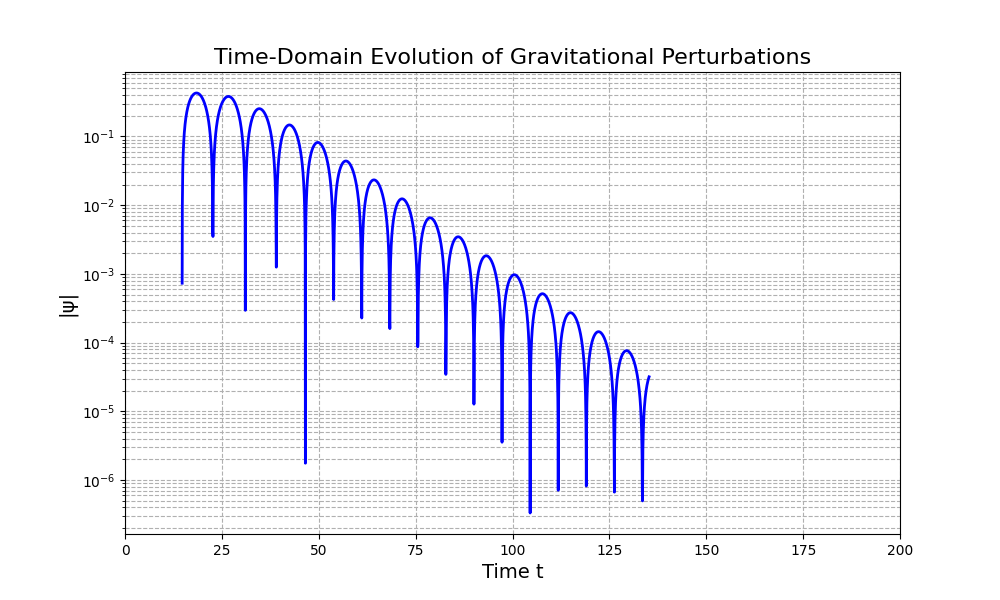}
         \caption{Time domain profile for gravitational perturbations}
         \label{fig:td_grav}
     \end{subfigure}
     \hfill
     \begin{subfigure}[t]{0.49\textwidth}
         \centering
         \includegraphics[width=\textwidth]{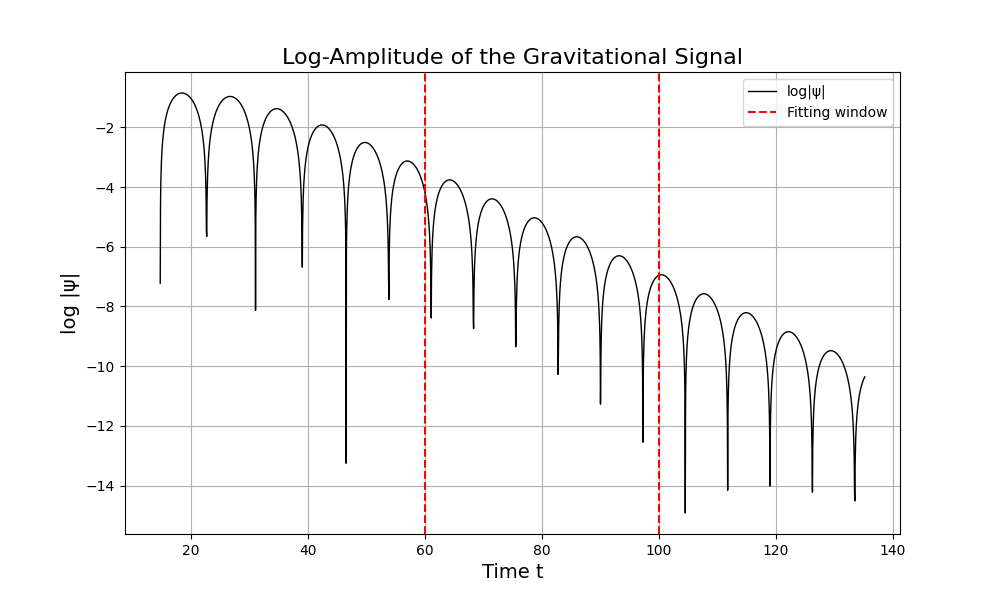}
         \caption{Log-amplitude and fitting window}
         \label{fig:td_grav_logamp}
     \end{subfigure}
     \begin{subfigure}[t]{0.49\textwidth}
         \centering
         \includegraphics[width=\textwidth]{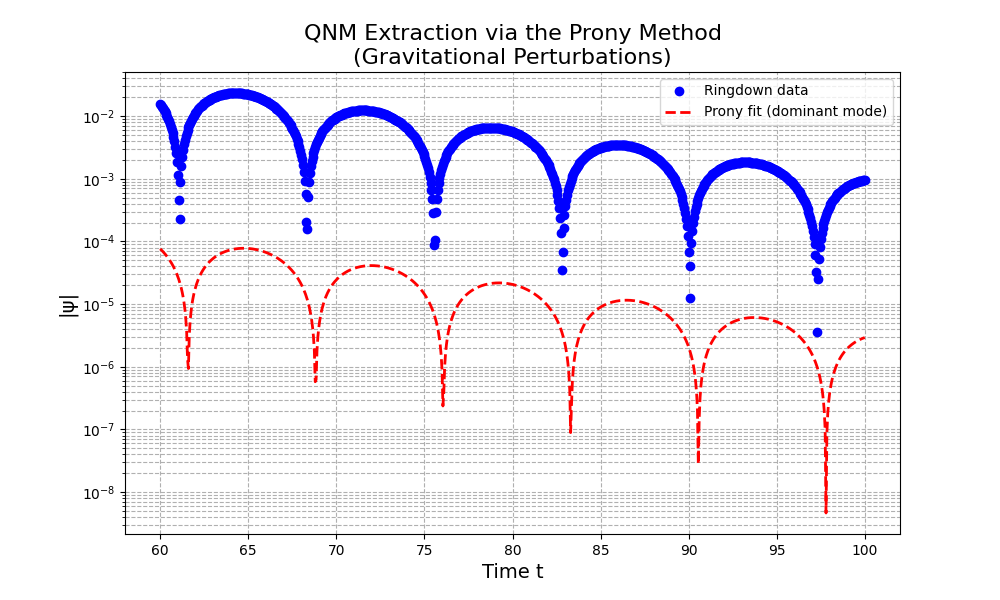}
         \caption{Prony fit of the time domain profile}
         \label{fig:prony_grav}
     \end{subfigure}
    \caption{Time domain profiles and Prony extraction for EM perturbations}
    \label{fig:td_grav_full}
\end{figure}

In this \textit{double-null} scheme, we specify appropriate initial data (Gaussian in this case) and find the time-domain profile data $\{\Phi(t=t_0),\Phi(t=t_0+h),\Phi(t=t_0+2h),\ldots\}$. We start by visualizing the time-domain profile for scalar perturbations for the $l=2$ mode with $\ell = -0.01$ and $\eta = 0.3$. An initial Gaussian pulse centered at $v_c = 20$ with width $\sigma = 2.5$ is used. In the numerical scheme, we appropriately handle the coordinate transformation ($r \rightarrow r_*$ and vice versa) and extract the signal at $r = 8$ ($r_* \sim 14.75$) for reference. The resulting time-domain profile is shown in Fig. \ref{fig:td_scalar}. Similarly, using the effective potentials for EM and gravitational perturbations in the same parameter region, we evaluated the corresponding time-domain profiles using the GPP method, as shown in Figs. \ref{fig:td_em} and \ref{fig:td_grav}. As discussed in the next section, we extract the corresponding QNMs from the time-domain signals with reasonable accuracy, and the behavior of the profiles with $\ell$ and $\eta$ are expected to retain the same dependence as that in the frequency-domain calculations.
% \begin{figure}
%     \centering
%     \includegraphics[width=0.7\linewidth]{TD_scalar.png}
%     \caption{Time-domain profile for scalar perturbation of the KR BH.}
%     \label{fig:td_scalar}
% \end{figure}
% \begin{figure}[htbp]
%     \centering
%     \includegraphics[width=0.7\linewidth]{KRBH_Prony.png}
%     \caption{Caption}
%     \label{fig:enter-label}
% \end{figure}
\subsection{Frequency extraction using the Prony method}
For the ringdown (QNM-dominated) portion of the signal, the Prony method approximates the signal by a sum of exponentially damped sinusoids \cite{berti2007mining}, one form of which can be written as:
\begin{equation}
  \psi(t)=\sum_{k} A_k \,e^{-i\omega_k t},
\end{equation}
with complex frequencies $\omega_k=\omega_{R,k}-i\,\gamma_k$ (where $\gamma_k>0$ for damped modes). For a sampling time $\Delta t$ in the discrete time domain, we have
\begin{equation}
  z_k=\exp(-i\omega_k\Delta t).
\end{equation}
The Prony method assumes a linear recurrence among the data and solves a linear least-squares problem to determine a polynomial whose roots $z_k$ yield the QNM frequencies via
\begin{equation}
  \omega_k=\frac{i}{\Delta t}\ln(z_k).
\end{equation}
Then, physical modes can be identified by requiring $\omega_{R,k}>0$ (oscillatory behavior) and $\gamma_k>0$ (damping, corresponding to $\operatorname{Im}(\omega_k)<0$). Fig. \ref{fig:td_logamp} shows the fitting window of the time-domain profile for scalar perturbations and Fig. \ref{fig:prony_scalar} shows the Prony fit to the time-domain signal. The estimated dominant QNM for the scalar case using the Prony method is $\omega_{0,2} = 0.487264 - 0.089473 i$, and that estimated using the Pad\'{e}-averaged WKB method is $\omega_{0,2} = 0.476361\, -0.0948474 i$. Thus, the QNM frequency is extracted with reasonable accuracy. For the EM case, the Prony-extracted dominant QNM is $\omega_{0,2} = 0.461418 - 0.088815 \,i$, and that using the Pad\'{e}-averaged WKB method is $0.45096\, -0.0930865 i$. For the gravitational case, the Prony-extracted frequency is $\omega_{0,2} = 0.433645 - 0.087799\, i$ and that extracted using the Pad\'{e}-averaged WKB method is $\omega_{0,2} = 0.423823\, -0.0911131 i$.

\section{Greybody Factors and Sparsity of Hawking Radiation}
\subsection{Greybody Factors}
\label{sec:gbfs}
In this section, we discuss the bounds on the GBFs. Drawing from the insights on the influence of the model parameters on the QNMs presented in the previous sections, we the GBFS associated with scalar perturbations and explore the impact of the model parameters on these bounds through an analytical approach. Analytical techniques for forecasting rigorous bounds on GBFs were first devised by Visser \cite{Visser:1998ke} and then improved by Boonserm \textit{et al.} \cite{Boonserm:2008zg}. More detailed investigations of these bounds have been reported by Boonserm \textit{et al.}, Yang \textit{et al.} \cite{Yang:2022ifo}, Gray \textit{et al.} \cite{Gray:2015xig}, Ngampitipan \textit{et al.} \cite{Ngampitipan:2012dq}, and others \cite{Chowdhury:2020bdi,Miao:2017jtr,Liu:2021xfs,Barman:2019vst,Xu:2019krv,Boonserm:2017qcq}. In the present case, we extend this investigation by analyzing a BH solution influenced by a global monopole within a self-interacting KR field, thereby enhancing our insight into GBFs in a different context.

\begin{figure}[!htbp]
     \centering
     \begin{subfigure}[t]{0.49\columnwidth}
         \centering
         \includegraphics[width=\textwidth]{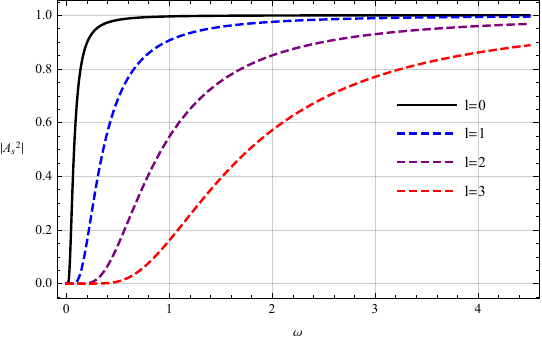}
         \caption{Variation with $l$}
         \label{GF1a}
     \end{subfigure}
     %\hfill
     \begin{subfigure}[t]{0.49\columnwidth}
         \centering
         \includegraphics[width=\textwidth]{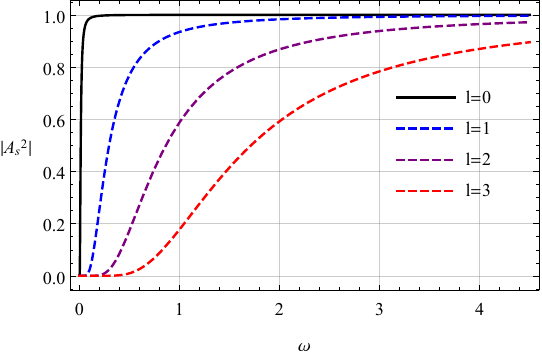}
         \caption{Variation with $l$}
         \label{GF1b}
     \end{subfigure}
    \caption{Rigorous bounds on greybody factors using $M = 1, \eta=0.9$. On the left panel, $\ell=0.1$ and on the right panel $\ell=-0.1$.}
    \label{GF1}
\end{figure}
\begin{figure}[!htbp]
     \centering
     \begin{subfigure}[t]{0.49\columnwidth}
         \centering         \includegraphics[width=\textwidth]{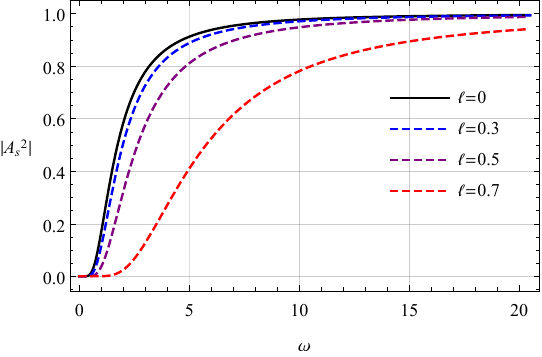}
         \caption{Variation with $\ell$}
         \label{GF2a}
     \end{subfigure}
     %\hfill
     \begin{subfigure}[t]{0.5\columnwidth}
         \centering
        \includegraphics[width=\textwidth]{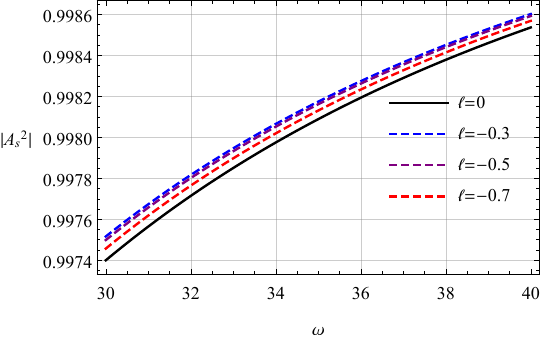}
         \caption{Variation with $\ell$}
         \label{GF2b}
     \end{subfigure}
    \caption{Rigorous bounds on greybody factors using $M = 1, \eta=0.9$. On the left panel, $\ell\geq0.1$ and on the right panel $\ell\leq-0.1$.}
    \label{GF2}
\end{figure}
\begin{figure}[!htbp]
     \centering
     \begin{subfigure}[t]{0.49\columnwidth}
         \centering         \includegraphics[width=\textwidth]{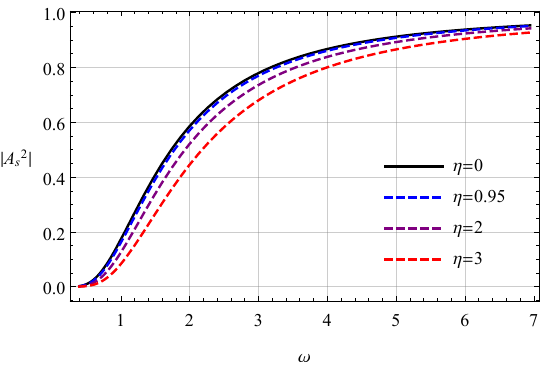}
         \caption{Variation with $\eta$}
         \label{GF3a}
     \end{subfigure}
     %\hfill
     \begin{subfigure}[t]{0.49\columnwidth}
         \centering
    \includegraphics[width=\textwidth]{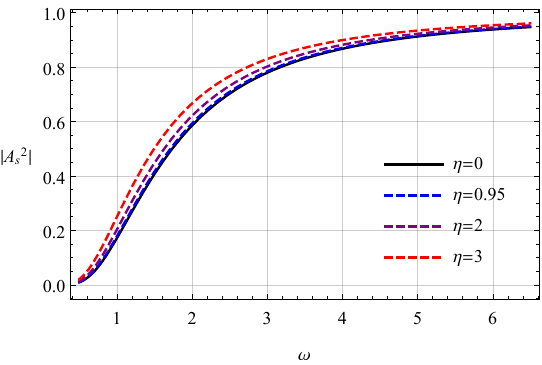}
         \caption{Variation with $\eta$}
         \label{GF3b}
     \end{subfigure}
    \caption{Rigorous bounds on greybody factors using $M = 1, l=3$. On the left panel, $\ell=0.1$ and on the right panel $\ell=-0.1$.}
    \label{GF3}
\end{figure}

%% Repeated statement, omitted. %%%We delve further into the constraints on greybody factors for black holes influenced by a global monopole within Kalb-Ramond field, confining it to massless scalar perturbations. 
We consider the Klein-Gordon equation of the massless scalar field, as described in the previous section. The reduced effective potential, $V_{eff}(r)$, is given by:
\begin{equation}
V_{eff}(r) = \frac{l(l + 1)A(r)}{r^{2}} + \frac{A(r)
A'(r)}{r}.\label{poten}
\end{equation}

We use the effective potential to investigate the lower bound of the GBFs in our BH solution. Following the work of Visser \cite{Visser:1998ke} and Boonserm \cite{Boonserm:2008zg}, a suitable method for determining this strict bound is provided by
\begin{equation} \label{bound}
A_g^2 \geq \operatorname{sech}^{2}\left(\frac{1}{2 \omega} \int_{-\infty}^{\infty}\left|V\right| \frac{d r}{A(r)} \right),
\end{equation}
where $A_g^2$ denotes the transmission coefficient.

Moreover, to account for the cosmological horizon-like effect as a term of the metric function, we adjust the boundary conditions as described by Boonserm \textit{et al.} \cite{Boonserm:2019mon}. The modified boundary condition is given by:
\begin{equation}
A_g^2 \geq A^2_{s}=\operatorname{sech}^{2}\left(\frac{1}{2 \omega} \int_{r_{H}}^{R_{H}} \frac{|V_{eff}|}{A(r)} d r\right)=\operatorname{sech}^{2}\left(\frac{A_{l}}{2 \omega}\right),
\end{equation}
where we define
 \begin{equation}
A_{l}=\int_{r_{H}}^{R_{H}} \frac{|V|}{A(r)} d r=\int_{r_{H}}^{R_{H}}\left|\frac{l(l+1)}{r^{2}}+\frac{
A^{\prime}(r)}{r}\right| d r.
\end{equation}

Here, $r_H$ is the event horizon and $R_H$ is the cosmological horizon of the BH. This specification provides a rigorous lower bound on the GBFs corresponding to the BH solution.

The behavior of the greybody bounds as a function of the model parameter is presented graphically in Figs. \ref{GF1}-\ref{GF3}. In the range $\ell^-\cup\ell^+$, we investigate how a change of sign in $\ell$ affects the corresponding behavior of the GBFs. We also study the variation with the global monopole charge and multipole moment $l$. It can be seen that the variation of the greybody bound with respect to $l$ for $\ell <0$ or $\ell >0$ remains unchanged (see Figs. \ref{GF1a}-\ref{GF1b}). In turn, the variation of the KR parameter $\ell$ spanned out across $\ell^-\cup\ell^+$ presents a detailed behavior of the greybody bound (see Figs. \ref{GF2a}-\ref{GF2b}). In this regime, the greybody boundary increases with decreasing $\ell$ in the $\ell^+$ range (\ref{GF2a}). By contrast, the increasing variation of the KR parameter $\ell$ in $\ell^-$ space leads to a decrease in the greybody bound (\ref{GF2b}). Figs. \ref{GF3a}-\ref{GF3b} highlight the effect of the global monopole charge $\eta$ on the greybody bound. For an increasing value of the KR field with $\ell=+0.1$ (Fig. \ref{GF3a}), the greybody bound tends to decrease, while for $\ell=-0.1$ (Fig. \ref{GF3b}), the grey body bound increases. Overall, the parameters $\eta$ and $\ell$ affect the greybody behaviors in a non-trivial manner, as expected from the previous discussion of the scalar potential and corresponding QNMs.

\subsection{Sparsity of Hawking Radiation}\label{sec:spar}

This section presents analyses on the sparsity Hawking radiation from the KR BH. Typically, a BH behaves analogously to a blackbody, emitting particles at a temperature close to the surface gravity. However, the Hawking radiation flux differs significantly from standard blackbody radiation in that it appears exceedingly sparse during the evaporation process. Sparsity is described as the mean time elapsed between the emission of consecutive quanta on time scales dictated by the energies of these quanta. It is defined as~\cite{Page:1976df,Gray:2015pma,Sekhmani:2024dyh,Sekhmani:2024fjn}
\begin{figure}[!htbp]
     \centering
     \begin{subfigure}[t]{0.49\columnwidth}
         \centering
    \includegraphics[width=\textwidth]{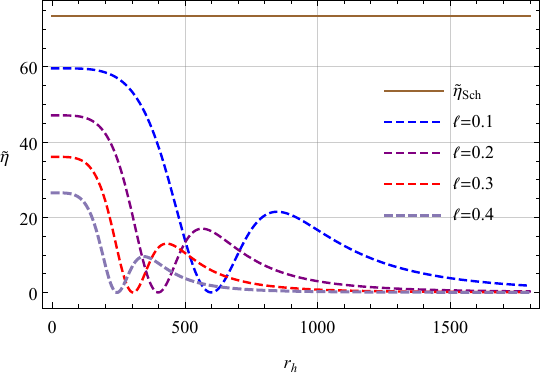}
         \caption{Variation with $\ell$}
         \label{ssp1a}
     \end{subfigure}
     %\hfill
     \begin{subfigure}[t]{0.49\columnwidth}
         \centering
    \includegraphics[width=\textwidth]{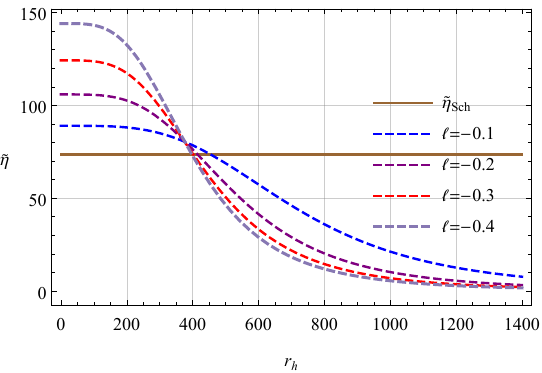}
         \caption{Variation with $\ell$}
         \label{ssp1b}
     \end{subfigure}
    \caption{Sparsity of Hawking radiation against $r_h$ using $\eta=0.01$. On the left panel, $\ell\geq0.1$ and on the right panel $\ell\leq-0.1$.}
    \label{ssp1}
\end{figure}
\begin{figure}[!htbp]
     \centering
     \begin{subfigure}[t]{0.49\columnwidth}
         \centering
\includegraphics[width=\textwidth]{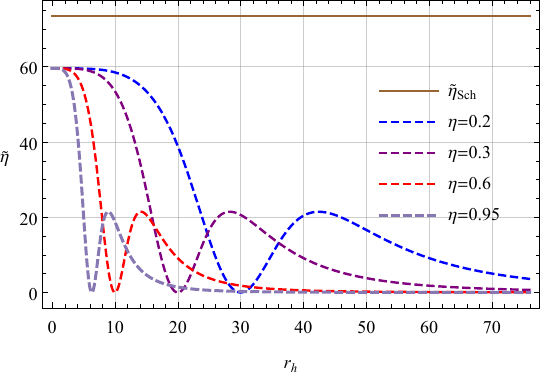}
         \caption{Variation with $\eta$}
         \label{ssp2a}
     \end{subfigure}
     %\hfill
     \begin{subfigure}[t]{0.49\columnwidth}
         \centering
    \includegraphics[width=\textwidth]{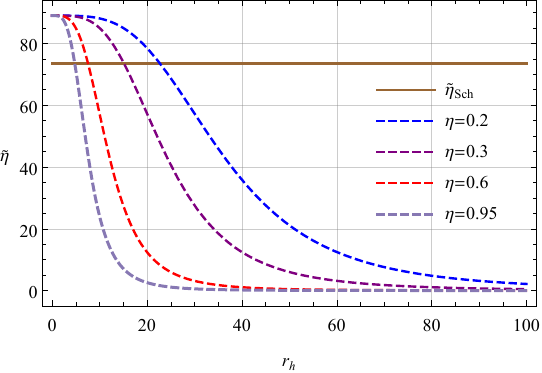}
         \caption{Variation with $\eta$}
         \label{ssp2b}
     \end{subfigure}
    \caption{Sparsity of Hawking radiation against $r_h$. On the left panel, $\ell=0.1$ and on the right panel $\ell=-0.1$.}
    \label{ssp2}
\end{figure}
\begin{equation}
\label{defSpars}
    \Tilde{\eta} =\frac{\mathcal{C}}{\Tilde{g} }\left(\frac{\lambda_t^2}{\mathcal{A}_{eff}}\right),
\end{equation}
 where $\lambda_t=2\pi/T$ denotes the thermal wavelength, $\mathcal{A}_{eff}=27 \mathcal{A}_{BH}/4$ acts as the effective surface area of the BH, $\mathcal{C}$ is a dimensionless constant, and $\Tilde{g}$ represents the spin degeneracy factor of the emitted quanta. Thus, $\lambda_t=8\pi r_h^2\,\Longrightarrow\,\eta_{Sch}=64\pi^3/27\approx73.49$ is obtained for the simple case of a Schwarzschild BH and spin-1 bosons with no emitted mass. Note that $\eta\ll 1$ for black body radiation under comparison.

To analyze the sparsity behavior of Hawking radiation, it is useful to look at the corresponding Hawking temperature, which is defined as follows: 
\begin{equation}
    T=\frac{1}{4\pi}\left(\frac{\mathrm{d}A(r)}{\mathrm{d}r}\right)_{r=r_h}
\end{equation}
from which the sparsity of Hawking radiation in terms of the parameter space can be expressed according to the following form:
\begin{equation}\label{spp}
   \Tilde{\eta}=\frac{256 \pi ^3 (\ell-1)^4 \left(\ell \left(\eta ^2 r^2+4\right)-4\right)^2}{3 \left(\eta ^4 \ell^2 r^4+6 \eta ^2 (\ell-1) \ell r^2+24 (\ell-1)^2\right)^2}.
\end{equation}
It is expected that the model parameters would affect the behavior of the sparsity. Specifically, the choice of the partial set $(\ell=0,\, \eta=0)$ allows us to probe a reduced form of the relevant sparsity such that
\begin{equation}
  \Tilde{\eta}(\ell=0,\, \eta=0)=\frac{64 \pi ^2 }{27},
\end{equation}
in which the corresponding sparsity across this limit matches that of the Schwarzschild case exactly. In addition, a closer examination shows that the corresponding sparsity \eqref{spp} coincides with the Schwarzschild one at
\begin{equation}
    r_h^c=\sqrt{\frac{(\ell-1) \left(\sqrt{3} \sqrt{\ell (3 \ell-8)}-3 \ell\right)}{\eta ^2 \ell}}
\end{equation}
where $\ell\leq-0.1$.

To unveil some of the scenarios characteristic of the study of sparsity in Hawking radiation, we need to examine the center the related behavior such that
\begin{align}
    \lim\limits_{r_h\to 0} \Tilde{\eta} &\approx \frac{64}{27} \pi ^3 (\ell-1)^2,
\end{align}
where the KR parameter $\ell$ affects the sparsity behavior at the center, generating three possible scenarios; the first scenario implies that the corresponding sparsity is exactly that of the Schwarzschild spacetime when $\ell=0$, the second scenario implies that $\Tilde{\eta}<\Tilde{\eta}_{Sch}$ when $\ell>0$, and the third serves to present the desired observation $\Tilde{\eta}>\Tilde{\eta}_{Sch}$ when $\ell<0$. By contrast, the large-distance sparsity of Hawking radiation provides
\begin{align}
    \lim\limits_{r_h\to \infty}\Tilde{\eta}&\approx0,
\end{align}
which indicates that $\Tilde{\eta}$ decreases monotonically and asymptotically approaches zero.

To interpret the behavior of the corresponding sparsity graphically, Figs. \ref{ssp1}-\ref{ssp2} describe the appropriate variation of the sparsity $\Tilde{\eta}$ as a function of the event horizon $r_h$. In this respect, two scenarios are proposed; specifically, $\ell<0$ (Figs. \ref{ssp1b} and \ref{ssp2b}) and $\ell>0$ (Figs. \ref{ssp1a} and \ref{ssp2a}). First, the scenario with $\ell>0$ is significantly more similar for both variations of $\ell$ and $\eta$, which depicts that for sufficiently small $r_h$, the sparsity lies below the standard value $\eta_{Sch}$, indicating that the emitted radiation is less sparse (more sparse for $\ell<0$) than the Hawking radiation at this evaporation stage. This is in contrast to the $\ell<0$ scenario, entailing at small $r_h$, the relation $\Tilde{\eta}>\Tilde{\eta}_{Sch}$ (Figs. \ref{ssp1b} and \ref{ssp2b}). As the event horizon radius increases, $\Tilde{\eta}$ falls monotonically and asymptotically approaches zero. In this state, the behavior appears more similar to black body radiation. This behavior can be observed for both $\ell>0$ and $\ell<0$. It is worth noting that the KR and global monopole parameters affect the decay rate in non-trivial manners. In contrast to the Schwarzschild BH, the sparsity decreases with increasing $r_h$.

\section{Discussions}
\label{sec:discn}
The comprehensive analyses in the previous sections of the paper aid in understanding the effects of the Lorentz-violating parameter, and more importantly in this regard, the presence of a global monopole charge on the dynamics of spherically symmetric BHs in KR gravity. Since solar system tests indicate that these parameters should be very small, investigating their nuanced effects on physical properties is of significant interest. To this end, our studies on perturbations of the KR BH with a global monopole charge establish the stability of the BH against perturbations. The analyses show that the QNMs are more sensitive to $\ell$ than to $\eta$; however, both parameters have distinguishable effects on the QNMs. A comparison of the estimated data with previously reported results for the static non-minimally coupled KR BH \cite{baruah2023quasinormal} shows that the estimated frequencies in this case are lower for all three types of perturbations. Moreover, differences from the GR-Schwarzschild case are subtle, potentially indicating difficulty in possible detection by near-future detectors. The QNMs are estimated with high accuracy using the Pad\'{e}-averaged WKB method, and the frequency and time-domain results exhibit reasonable agreement. However, it is noted that potential avenues for improvement in the numerical scheme developed for the time-domain analysis remain, including but not limited to using more accurate frequency-extraction approaches such as the matrix-pencil method. In a future study, we aim to obtain more meaningful inferences from BH spectroscopic studies of rotating KR BHs.
%%% edited for conciseness %%%

Next, the analyses of the effects of the global monopole charge on the greybody bounds yields fascinating results. On separating the spectrum of the KR parameter $\ell$ into $\ell^-\cup \ell^+$, two different regimes emerged. In particular, analysis in the $\ell^-$ space showed opposite results to that in the $\ell^+$ space. In addition, the results showed that the effect of the multipole moment $l$ remains unchanged when transitioning from the $\ell^-$ to the $\ell^+$ set, implying that the multipole moment no longer interacts with the KR parameter $\ell$. In contrast, the variation of the greybody bound is affected by the variation of the global monopole charge in the $\ell^-$ to $\ell^+$ transition.

Lastly, the sparsity of Hawking radiation in light of the influence of the global monopole on the BH in the self-interacting KR field was inspected. The analysis has provided valuable and practical insight into the parameter space. By dividing the possible interval of the KR parameter $\ell$ into $\ell^-\cup\ell^+$, the results are interpreted from two different points of view. Depending on the positivity (negativity) of $\ell$ at sufficiently small $r_h$, the sparsity is found to be lower (higher) than the standard value $\eta_{Sch}$, which means that the radiation emitted is less (more) sparse than the Hawking radiation at this stage of evaporation. It was observed that as the radius of the event horizon increases, $\Tilde{\eta}$ decreases monotonically and asymptotically approaches zero. In this state, the behavior approximately resembles black body radiation.

\bibliographystyle{JHEP}
\bibliography{biblio.bib}
\end{document}